\documentclass{elsart}
\usepackage[applemac]{inputenc}
\usepackage{graphicx}
\usepackage{amsmath}
\usepackage{times}
\usepackage{txfonts}
\usepackage{array}
\usepackage{natbib}
\usepackage{rotating}
\usepackage[dvipsnames,usenames]{color}
\usepackage{setspace}
\usepackage{setspace}

\newcommand{\te}[1]{10^{#1}}
\newcommand{\ut}[1]{\hspace{1mm}\mathrm{#1}}
\newcommand{\vc}[1]{\boldsymbol{\mathrm #1}}
\newcommand{\vecu}{\vc{u}} 
\newcommand{\vecB}{\vc{B}}

\newcommand{\ddp}[2]{\displaystyle \dfrac{\displaystyle \partial #1}{\displaystyle \partial #2}}

\newcommand{\cit}[1]{\let\temp=\\ \centering #1\let\\=\temp}
\newcommand{\rev}[1]{{#1}}
\newcommand{\revm}[1]{{#1}}

\newcolumntype{L}[1]{>{\raggedright\let\newline\\\arraybackslash\hspace{0pt}}m{#1}}
\newcolumntype{C}[1]{>{\centering\let\newline\\\arraybackslash\hspace{0pt}}m{#1}}
\newcolumntype{R}[1]{>{\raggedleft\let\newline\\\arraybackslash\hspace{0pt}}m{#1}}

\begin{document}

\begin{frontmatter}
\title{Geomagnetic acceleration and rapid hydromagnetic wave dynamics in advanced numerical simulations of the geodynamo}

{\it Geophys. J. Int. 214, 531-547, 2018}

\author{Julien Aubert}
\address{Institut de Physique du Globe de Paris, Sorbonne Paris Cit\'e, Universit\'e Paris-Diderot, CNRS, 1 rue Jussieu, F-75005 Paris, France.}

\begin{abstract}
Geomagnetic secular acceleration, the second temporal derivative of Earth’s magnetic field, is a unique window on the dynamics taking place in Earth’s core. In this study, the behaviours of the secular acceleration and underlying core dynamics are examined in new numerical simulations of the geodynamo that are dynamically closer to Earth's core conditions than earlier models. These new models reside on a theoretical path in parameter space connecting the region where most classical models are found to the natural conditions. The typical time scale for geomagnetic acceleration is found to be invariant along this path, at a value close to 10 years that matches Earth's core estimates. Despite this invariance, the spatio-temporal properties of secular acceleration show significant variability along the path, with an asymptotic regime of rapid rotation reached after 30\% of this path (corresponding to a model Ekman number $E=3~\te{-7}$). In this regime, the energy of secular acceleration is entirely found at periods longer than that of planetary rotation, and the underlying flow acceleration patterns acquire a two-dimensional columnar structure representative of the rapid rotation limit. The spatial pattern of the secular acceleration at the core-mantle boundary shows significant localisation of energy within an equatorial belt. Rapid hydromagnetic wave dynamics is absent at the start of the path because of insufficient time scale separation with convective processes, weak forcing and excessive damping but can be clearly exhibited in the asymptotic regime. This study reports on ubiquitous axisymmetric geostrophic torsional waves of weak amplitude relatively to convective transport, and also stronger, laterally limited, quasi-geostrophic Alfvén waves propagating in the cylindrical radial direction from the tip of convective plumes towards the core-mantle boundary. In a system similar to Earth's core where the typical Alfvén velocity is significantly larger than the typical convective velocity, quasi-geostrophic Alfvén waves are shown to be an important carrier of flow acceleration to the core surface that links with the generation of strong, short-lived and intermittent equatorial pulses in the secular acceleration energy. The secular acceleration time scale is shown to be insensitive to magnetic signatures from torsional waves because of their weak amplitude, and from quasi-geostrophic Alfvén waves because of their intermittent character, and is therefore only indicative of convective transport phenomena that remain invariant along the parameter space path.
\end{abstract}

\begin{keyword}
Dynamo: theories and simulations; Rapid time variations; Magnetic anomalies: modelling and interpretation; Satellite magnetics. \end{keyword}

\end{frontmatter}

\section{\label{intro}Introduction}
The Earth's magnetic field of internal origin varies on time scales ranging from \rev{less than} a year to hundred million years. The first time derivative of the magnetic signal or geomagnetic secular variation (SV) enables the retrieval of the fluid flow at the top of Earth's core \citep[e.g.][]{Holme2015}. This in turn provides valuable information on the kinematics of the geodynamo process that generates the field, for instance concerning the current decay of the magnetic dipole \citep{Finlay2016} or the evolution of the South Atlantic intensity anomaly \citep{Aubert2015}. To go beyond a kinematic description and investigate the underlying dynamics, one needs to take the second derivative of the magnetic signal, known as the geomagnetic secular acceleration (SA). The classical method for mathematically estimating the SV and SA is the production of parameterised continuous descriptions of the geomagnetic field in space and over a period of time, known as geomagnetic field models. Early models such as {\it gufm1} \citep{Jackson2000} have been useful to constrain the SV from marine and ground observatory geomagnetic data over the past four hundred years, but penalised the SA to reduce the model complexity in time. Subsequent geomagnetic field models such as the \rev{CM \citep{Sabaka2015}}, CHAOS \citep{Olsen2006,Finlay2016b}, GRIMM \citep{Lesur2010}, C3FM \citep{Wardinski2012} and COV-OBS \citep{Gillet2013,Gillet2015b} series use the wealth of data provided by \rev{low-Earth orbiting satellite missions} in addition to ground observatory data, together with \rev{a variety of distinct strategies to achieve} weaker and refined temporal regularisations. This has enabled drastic improvements in the description of the SA \citep{Olsen2006}. The latest generation of models making use of data from the {\it Swarm} constellation \citep[e.g.][]{Finlay2016b} provide a continuous description of the geomagnetic SA over the past two decades, with horizontal resolution of approximately 2000 km at the core surface, corresponding to degree $\ell=9$ of a spherical harmonic expansion, and temporal resolution down to about a year on the largest length scales. 

One of the most important results from high-resolution satellite geomagnetic field models is how they render geomagnetic jerks, or abrupt discontinuities in the SA of internal origin that have been initially identified in earlier ground observatory records \citep[see e.g.][]{Mandea2010}. The 2003 and 2007 jerks have been related to the existence of a short-lived, intense pulse in the SA energy at the core-mantle boundary \citep{Chulliat2010}, occurring around 2006 close to the equator. \rev{More recently, similar pulses have been found to occur approximately every three years, accounting for additional jerks around 2011 and 2014 \citep{Chulliat2014,Chulliat2015,Torta2015, Soloviev2017}. These observations raise} concerns about the potential limit of the current approach through which the SA is determined. One can indeed question whether the currently available images of the pulses indeed represent the real phenomenon in its full detail or whether they correspond to an energy stacking artifact related to temporal subsampling of an even richer signal, the high-frequency content of which is not accessible because of the masking of internal signals by rapid contributions of external origin. Reproduction of the SA pulses in self-consistent numerical simulations of the geodynamo has so far been elusive, but would be a valuable tool to constrain the temporal content of pulses and understand their dynamical mechanism. Significant geomagnetic SA also exists at higher latitudes in the period 2000-2016, beneath a region extending from Siberia to Canada in the Eastern hemisphere \citep{Finlay2016b}. Recent core-surface flow inversions from the geomagnetic SV have suggested the existence of a planetary-scale, eccentric, axially columnar gyre in the outer core \citep{Pais2008,Gillet2013,Aubert2013a,Aubert2014,Gillet2015,Pais2015}, and the high-latitude SA signal can be interpreted as a localised acceleration of this gyre \citep{Livermore2017}. Other interesting phenomena related to SA can also be investigated prior to 1997, but due to a considerable loss of spatial and temporal resolution, only decadal changes of global features can be analysed. A sudden acceleration of the North \rev{magnetic} pole motion has for instance been observed between 1989 and 2002 \citep{Chulliat2010b}, which has been ascribed to an event of polar magnetic flux expulsion. 

Improvements in the resolution and temporal extent of geomagnetic field models have also led to the determination of the typical secular acceleration time scales $\tau_{\mathrm{SA}} (\ell)$ from the ratio of the energy present at each spherical harmonic degree $\ell$ in the SV and SA. Several reports from different geomagnetic field models have hinted at an approximately constant $\tau_{\mathrm{SA}}(\ell) \approx 10 \ut{yr}$ at all resolvable time scales \citep{Holme2011,Christensen2012}. In the latter study this constant time scale was interpreted as reflecting the action of fluid flow acceleration on the magnetic field (as opposed to the action of fluid flow on the secular variation, another source for secular acceleration, see section \ref{breakdown}). In Earth's core, it is expected that flow acceleration should feature a rich variety of rapid (i.e. time scales much shorter than that of convection) hydromagnetic wave phenomena in addition to the signature of the comparatively slower convection. Given that magnetic torsional waves in Earth's core have been identified with a time scale of about 6 years \citep{Gillet2010}, one could be tempted to ascribe the estimated $\tau_{\mathrm{SA}}(\ell) \approx 10 \ut{yr}$ to such waves. \cite{Christensen2012} also showed that both the shape and the actual value of the $\tau_{\mathrm{SA}} (\ell)$ spectrum could be matched in self-consistent numerical simulations of the convective geodynamo, provided that the ratio between the magnetic diffusion time $\tau_{\eta}$ and core overturn time $\tau_{U}$ (the magnetic Reynolds number) has an Earth-like value of about 1000. Due to computational limitations, the numerical dynamos used by \cite{Christensen2012} however featured a modest level of separation between $\tau_{\eta}$, $\tau_{U}$ on the one hand, and two other key time scales, the Alfvén time $\tau_{A}$ and rotational time $\tau_{\Omega}$ (see section \ref{param} for definitions). This implies that hydromagnetic waves of any origin are absent because of weak forcing, strong damping, and because their typical time scale $\tau_{A}$ tends to overlap with the time scale $\tau_{U}$ of convection \citep[e.g.][]{Wicht2010}. For this reason, the possibility to fully reproduce the $\tau_{\mathrm{SA}} (\ell)$ spectrum came as a surprising result and suggested that contrary to the above expectations, the processes responsible for flow acceleration in Earth's core could in fact be similar to those, entirely driven by convection, already accounted for in the standard dynamo simulations. 

Numerical geodynamo simulations achieving significantly more realistic levels of separation between $\tau_{\eta}$, $\tau_{U}$, $\tau_{A}$ and $\tau_{\Omega}$ have gradually appeared in the recent litterature \citep{Wicht2010,Teed2014,Teed2015,Schaeffer2017}, and it has been shown that the rapid torsional waves missing in the standard models are indeed featured in these new models. Systematic analysis of the SA in such models could give an opportunity to clearly highlight the signature of rapid dynamics, guide the interpretation of observations, better assess their quality level, and solve the above conundrum concerning the SA time scale. Achieving high temporal time scale separation however comes at a high computational cost, thereby limiting the possibility to do systematic parameter space sampling in the numerical simulations. To circumvent this issue, an alternative approach has been proposed \citep[][from hereafter A17]{Aubert2017}, based on the formulation of a unidimensional theoretical path in parameter space that connects the classical numerical dynamo simulations such as used in \cite{Christensen2012} to Earth's core conditions. Large-eddy simulations of reasonable accuracy could be conducted over half of this parameter space path, leading to models that can be considered as the closest currently available to the physical conditions of Earth's core. One of the striking properties of these simulations is that they feature kinematic invariance, in the sense that the magnetic field, flow and density anomaly morphologies remain similar at large scale along the path. In contrast, the dynamics is not expected to be invariant, because the asymptotic regime of rapid rotation, strong magnetic control on the flow (or strong-field dynamo action), and low attenuation is gradually enforced as simulations progress along the path. This strengthens the prospect of observing rich hydromagnetic wave dynamics in addition to convection in the most advanced models. The purpose of this study is to systematically investigate the spatial and temporal SA properties from the path simulations of A17, highlight the new underlying dynamics, and examine their observable signatures. The study is organised as follows. Section \ref{model} presents the numerical dynamo model and methods. Results are presented in section \ref{results} and discussed in section \ref{discu}. 

\section{\label{model}Model and methods.}

\subsection{Numerical geodynamo model set-up.}
The magnetohydrodynamic equations and numerical set-up correspond to the configuration CE \rev{(for Coupled Earth model)} previously introduced in A17, that are recalled here within a geophysical context. The model solves for Boussinesq convection, thermochemical density anomaly transport and magnetic induction in the magnetohydrodynamic approximation in an electrically conducting and rotating spherical fluid shell representing the outer core, with a rotation rate $\Omega$, density $\rho$ and thickness $D=r_{o}-r_{i}$ such that $r_{i}/r_{o}=0.35$ as in the Earth. The unknowns are the velocity field $\vecu$, the magnetic field $\vecB$ and the density anomaly field $C$. An important prerequisite to the geophysical analysis of SV and SA is the reproduction of the geomagnetic westward drift. This is achieved here as in \cite{Aubert2013b,Pichon2016}, by modelling couplings at the global scale between the outer core, inner core and mantle, such that indirect coupling between the outer core and mantle via the inner core dominates direct coupling through the core-mantle boundary. To this end, the outer core is electromagnetically coupled to a solid inner core of radius $r_{i}$, itself gravitationally coupled to a solid outer shell representing the mantle between radii $r_{o}$ and $1.83 r_{o}$. Both the inner core and mantle feature a time-dependent axial differential rotation with respect to the outer core, with respective angular velocities $\Omega_{ic}$ and $\Omega_{m}$, and the gravitational torque felt by the inner core is defined as $\Gamma=-\xi~\rho D^{5} \Omega (\Omega_{ic}-\Omega_{m})$ with $\xi=0.75$ in all simulations. This indirect coupling co-exists with a direct electromagnetic coupling between the outer core and mantle. The three regions (inner core, outer core, mantle) are assigned moments of inertia respecting the proportions relevant to the Earth, and the ensemble has a constant angular momentum defining the planetary rotation rate $\Omega$. Electrically conducting boundary conditions are used at both fluid shell boundaries \rev{\citep[see][]{Pichon2016}}. The electrical conductivity of the inner core is set at the same value $\sigma_c$ as that of the outer core. The mantle features an electrically conducting region at its base, with thickness $\Delta$ and conductivity $\sigma_{m}$. In the model the dimensionless conductance has been set to a value $\Delta\sigma_{m}/D\sigma_{c}=\te{-4}$ that is at the mid-point of current geophysical estimates \citep{Pichon2016}. Stress-free mechanical boundary conditions are used at both boundaries. The model operates at low viscosity regimes where stress-free and rigid boundary conditions become undistinguishable (A17), but the use of the former alleviates the computational cost by removing the need to numerically resolve the viscous boundary layers. 

The conditions for thermochemical density anomaly are of fixed-flux type at both boundaries. This configuration is geophysically appropriate because the core-mantle boundary heat flow and the density anomaly flux at the inner core boundary are both controlled by the cooling rate of the surrounding mantle. Furthermore, it alleviates the computational cost by removing the need to resolve thermochemical boundary layers. Finally, to a good approximation the convective power becomes an input parameter in this configuration (A17). At the inner core boundary, an homogeneous density anomaly flux $F$ corresponding to a convective power input $P$ is prescribed, while the homogeneous density anomaly flux vanishes at the core-mantle boundary. A volumetric sink term is then present in the density anomaly transport equation to conserve mass. Within the Boussinesq framework, this model represents a geophysical situation with bottom-driven chemical convection originating from inner core solidification, a fully convective outer core and an exactly adiabatic heat flow $Q_{ad}$ at the core-mantle boundary. The geographical localisation of the SV is sensitive to possible lateral density-anomaly flux heterogeneities at both boundaries of the outer core, and it will be shown here that this also applies to the SA. The following setup favours the hypothesis of bottom-up, rather than top-down heterogeneous control on the SV and SA \citep{Aubert2013b}. At the inner core boundary, a longitudinally hemispheric density anomaly flux pattern is imposed on top of the homogeneous flux, meant to represent the effect of an asymmetric growth of the inner core. The maximum flux is located at longitude $90^{\circ}$E, and the peak-to-peak amplitude $\Delta f_{i}$ is such that $4\pi r_{i}^{2} \Delta f_{i}/ F=0.8$. At the core-mantle boundary, an heterogeneous flux pattern derived from lower mantle seismic tomography is imposed, meant to represent thermal control from the heterogeneous lower mantle. The morphology is the same as in \cite{Aubert2008_2,Aubert2013b}. To scale the amplitude, the model uses the geophysical estimates $\Delta q=40 \ut{mW/m^{2}}$ for the peak-to-peak value of heat flow heterogeneities at the core mantle boundary \citep{Lay2006,VanDerHilst2007}, and $q_{ad}=100 \ut{mW/m^{2}}$ for the homogeneous adiabatic heat flow per unit surface \citep[corresponding to an integral heat flow $Q_{ad}= 15 \ut{TW}$,][]{Pozzo2012}. The translation of these values to numerical dynamo parameters is not straightforward, because in the Boussinesq framework the homogeneous density anomaly flux vanishes at the core-mantle boundary. Consideration of core thermodynamics however helps to circumvent this issue, and leads to a relationship between $\Delta q/q_{ad}$ and the ratio $4\pi r_{i}^{2}\Delta f_{o}/F$ of the peak-to-peak core-mantle boundary density anomaly flux heterogeneity $\Delta f_{o}$ to imposed homogeneous flux $F$ at the inner core boundary \citep[see equation 5 in the methods of][]{Aubert2013b}. This relationship leads us to prescribe $4\pi r_{i}^2 \Delta f_{o}/F=0.029$ (or more simply $4\pi D^2 \Delta f_{o}/F=0.1$). 

\subsection{\label{param}Control parameters, path theory and key time scales.}
The four main control parameters of the model are the flux-based Rayleigh, Ekman, Prandtl and magnetic Prandtl numbers
\begin{eqnarray}
Ra_{F}&=&\dfrac{g_{o}F}{4\pi\rho\Omega^{3}D^{4}},\\
E&=&\dfrac{\nu}{\Omega D^{2}},\\
Pr&=&\dfrac{\nu}{\kappa},\\
Pm&=&\dfrac{\nu}{\eta}.
\end{eqnarray}
Here $g_{o}$, $\nu$, $\kappa$ and $\eta$ are respectively the gravity at the outer boundary  \rev{($r=r_{o}$)} of the model, the fluid viscosity, thermo-chemical and magnetic diffusivities. Note that $\eta=1/\mu \sigma_{c}$, where $\mu$ is the magnetic permeability of the fluid. The concept of a unidimensional path in parameter space has been recently introduced in A17, by showing that the variations in these control parameters that are necessary to bridge the gap between standard models such as the coupled Earth dynamo \citep{Aubert2013b} and Earth's core conditions can be represented as power laws of a single variable $\epsilon$ (the path parameter). Any model along the path is defined using the following rules:
\begin{eqnarray}
Ra_{F}&=&\epsilon Ra_{F} (CE),\\
E&=&\epsilon E (CE),\\
Pr&=&1,\\
Pm&=&\sqrt{\epsilon} Pm (CE).
\end{eqnarray}

\begin{table}
\setstretch{0.8}\small
\hspace*{-1.4cm}\begin{tabular}{lm{2cm}rrrrrr}
\hline\\[-0.5cm]
Label & Path position and type & $\epsilon$ & $E=\dfrac{\tau_{\Omega}}{\tau_{\nu}}$ & $E_\eta=\dfrac{\tau_{\Omega}}{\tau_{\eta}}$ & $Ro=\dfrac{\tau_{\Omega}}{\tau_{U}}$ & $A=\dfrac{\tau_{A}}{\tau_{U}}$ & $S=\dfrac{\tau_{A}^{-1}}{\tau_{\nu}^{-1}+\tau_{\eta}^{-1}}$ \\[0.4cm]
\hline
Start & 0\%  DNS & 1 & $3~\te{-5}$ & $1.2~\te{-5}$  & $1.11~\te{-2}$ & 0.70 & 380 \\
& 7\%  DNS & 0.33 & $\te{-5}$ & $6.94~\te{-6}$ & $7.19~\te{-3}$ & 0.61 & 700 \\
& 7\%  LES & 0.33 & $\te{-5}$ & $6.94~\te{-6}$ & $6.86~\te{-3}$ &  0.57 &   710    \\
&14\% DNS & 0.1 & $3~\te{-6}$ & $3.75~\te{-6}$ & $4.10~\te{-3}$ & 0.47 & 1300 \\
&14\% LES & 0.1 & $3~\te{-6}$ & $3.75~\te{-6}$ & $3.92~\te{-3}$ & 0.44 &  1300 \\
&21\% DNS & $3.33~\te{-2}$ & $\te{-6}$ & $2.22~\te{-6}$ & $2.35~\te{-3}$ & 0.35 & 2100 \\
&21\% LES & $3.33~\te{-2}$ & $\te{-6}$ & $2.22~\te{-6}$ & $2.30~\te{-3}$ & 0.34 & 2100 \\
&29\% LES & $\te{-2}$ & $3~\te{-7}$ & $1.2~\te{-6}$ & $1.26~\te{-3}$ & 0.24 & 3400 \\
&36\% LES & $3.33~\te{-3}$ & $\te{-7}$ & $6.94~\te{-7}$ & $7.42~\te{-4}$ & 0.19 & 4900 \\
&43\% LES & $\te{-3}$ & $3~\te{-8}$ & $3.80~\te{-7}$ & $4.17~\te{-4}$ & 0.15 & 6900 \\
Midpath & 50\% LES  & $3.33~\te{-4}$ &	 $\te{-8}$ & $2.22~\te{-7}$ & $2.40~\te{-4}$ & 0.11 & 9100\\
End  & 100\% Ext.  & $\te{-7}$ & $3~\te{-12}$ & $3.8~\te{-9}$ & $3.5~\te{-6}$ & 0.012 & 73600 \\
Earth  & & & $\mathcal{O}(\te{-15})$ & $\mathcal{O}(\te{-9})$ & $\approx 3~\te{-6}$ & $\approx 0.015$ & $\mathcal{O}(\te{5})$\\
\hline
\end{tabular}
\vspace*{0.1cm}\setstretch{1}\normalsize
\caption{\label{tsratios}Set of numerical models and ratios of key time scales (see text for definitions), together with scaling extrapolations obtained at the end of path and estimates at Earth's core conditions. Types DNS, LES and Ext. respectively refer to fully-resolved (direct) numerical simulations, large-eddy simulations, and scaling extrapolations, all obtained from data in A17, at the exception of the DNS at 21\% of the path which is original to the present work.}
\end{table}

Here $Ra_{F}(CE)=2.7~\te{-5}$, $E(CE)=3~\te{-5}$ and $Pm(CE)=2.5$ are the control parameters of the coupled Earth dynamo model defining the start of the path ($\epsilon=1$), and A17 have shown that conditions relevant to Earth's core are reached at the end of path defined by $\epsilon=\te{-7}$. The model cases (Table \ref{tsratios}) sample half-decades in $\epsilon$ from $\epsilon=1$ to $\epsilon=3.33~\te{-4}$, this latter value defining the model at the middle of the path, from herafter the Midpath model. The parameters of the Midpath model are the closest to Earth's core conditions employed to date in a numerical dynamo simulation, at the expense of a large-scale approximation (see section \ref{numerical}). Table \ref{tsratios} introduces a number of key time scales to illustrate the path approach: the rotational time $\tau_\Omega=1/\Omega$, convective overturn time $\tau_{U}=D/U$, Alfvén time $\tau_{A}=\sqrt{\rho \mu} D/B$, magnetic and viscous diffusion times $\tau_{\eta}=D^{2}/\eta$, $\tau_{\nu}=D^{2}/\nu$. Here $U$ and $B$ are respectively the root-mean-squared velocity and dynamo-generated magnetic field in the fluid shell. Four classical dimensionless time scale ratios are presented in addition to the Ekman number $E=\tau_{\Omega}/\tau_{\nu}$: the magnetic Ekman number $E_{\eta}=E/Pm=\tau_{\Omega}/\tau_{\eta}$, Rossby number $Ro=\tau_{\Omega}/\tau_{U}$, Alfvén number $A=\tau_{A}/\tau_{U}$ and Lundquist number $S=(\tau_{A}(\tau_{\nu}^{-1}+\tau_{\eta}^{-1}))^{-1}$ \citep[a measure of the attenuation of Alfvén waves,][]{Jault2008}. Throughout the path, the magnetic Reynolds number $Rm=\tau_{\eta}/\tau_{U}=Ro/E_{\eta}$ comparing the magnetic diffusion time to the overturn time remains roughly constant (a defining property of the path), at a geophysically reasonable value $Rm\approx 1,000$. As illustrated by the trends seen in Table \ref{tsratios}, the scaling analysis performed in A17 has shown that all time scale ratios can be expressed as power laws of $\epsilon$, the extrapolation of which to the end of path closely match the estimates for Earth's core, at the exception of the viscous time scale for which the limit $\tau_{\nu}\gg\tau_{\eta}$ (i.e. $Pm\ll 1$, inviscid behaviour) is considered sufficient. Numerical models taken along this path can therefore be understood as continuously progressing from imperfect towards geophysically appropriate conditions in all relevant aspects of their inputs and outputs.

\begin{figure}
\centerline{\includegraphics[width=10cm]{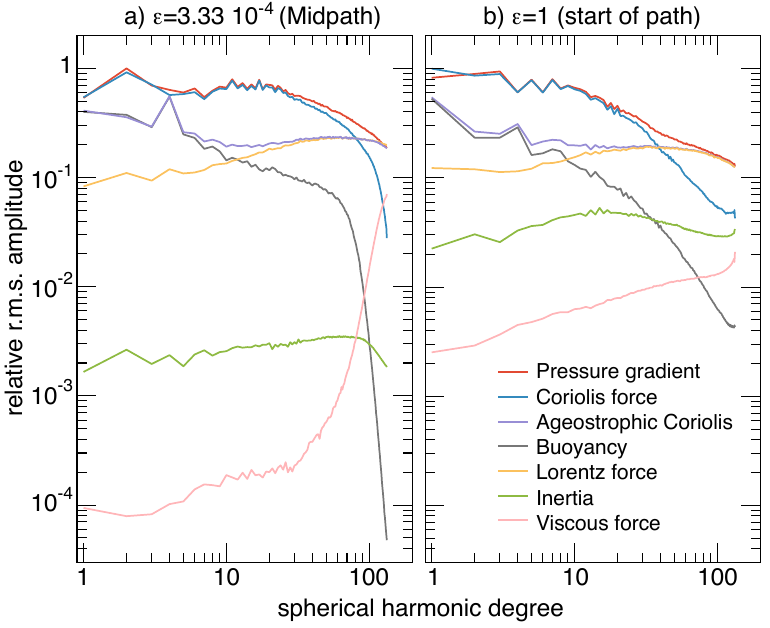}}
\caption{\label{forcebal} Force balance structure in snapshots from the Midpath (a) and \rev{start-of-path} (b) models, presented as the relative r.m.s. amplitude of each force at each spherical harmonic degree $\ell$. See A17 for technical computation details. \rev{Note also that the published version of A17 contains a legend mismatch in its figure 2, where the labels of the Coriolis and pressure forces are interverted. This mismatch is corrected here.}}
\end{figure}

The regime of Earth's core is an asymptotic limit of rapid rotation, strong convective power input and strong magnetic fields where $\tau_{\eta}\gg\tau_{U}\gg\tau_{A}\gg\tau_{\Omega}$ (Table \ref{tsratios}). The corresponding force balance structure (see A17) can be illustrated in a model attaining this limit, such as the Midpath model (Fig.  \ref{forcebal}a). A zeroth-order geostrophic balance is observed between the pressure and Coriolis forces, indicating strong rotational control on the dynamics. At first order, the so-called MAC balance is observed between the Lorentz, buoyancy and ageostrophic Coriolis forces. At second order, inertial forces are much smaller than the MAC forces, indicating the possibility of strong buoyant and magnetic forcing on the flow. Finally viscosity is negligible, despite the use of hyperdiffusivity which raises its strength at the smallest scales (though the viscous force does not perturb the MAC balance at any scale). The dynamical signature corresponding to the asymptotic regime and to this organisation of forces is the presence of rotationally-dominated, strongly forced, weakly attenuated magneto-inertial waves at a short typical time scale $\tau_{A}$ that co-exist with turbulent convection with a longer time-scale $\tau_{U}$ \cite[e.g.][]{Schaeffer2017}. In contrast, numerical models at the start of the path, similar to those used in \cite{Christensen2012}, are characterised by $\tau_{\eta}\gg\tau_{U}\approx\tau_{A}$, and a modest level of separation between ($\tau_{A}$,$\tau_{U}$) and $\tau_{\Omega}$. In this context, magneto-inertial waves are hindered \citep{Wicht2010} because of low forcing (see weak separation of inertia and MAC forces in Fig. \ref{forcebal}b), high damping (a measured by the low value of the Lundquist number $S$ in Table \ref{tsratios}), insufficient rotational control on the flow and insufficient time scale separation with convection. From this discussion, important qualitative changes can therefore be expected between the start of path and Midpath models regarding the properties of SA that reflect the system dynamics, with the Midpath model being representative of the dynamics taking place in Earth's core. 

\subsection{Secular variation, acceleration times and rescaling of dimensionless model output.}
\begin{table}
\begin{center}
\hspace*{-1.3cm}\setstretch{0.8}\small
\begin{tabular}{lm{2cm}rrrrrrr}
\hline\\[-0.5cm]
Label & Path position and type & $\dfrac{\tau_{\mathrm{SV}}^{1}}{\tau_{\eta}}$ & $\dfrac{\tau_{\mathrm{SA}}^{0}}{\tau_{\eta}}$ & $\tau_{\eta}$ (yr) & $\tau_{\mathrm{SA}}^{0}$ (yr)  & $\tau_{U}$ (yr) & $\tau_{A}$ (yr) & $2\pi\tau_{\Omega}$ (yr)\\[0.3cm]
\hline
Start & 0\% DNS & $3.34~\te{-3}$ & $9.25~\te{-5}$ &  $1.24~\te{5}$ & 11.5 & 134 & 93.0 & 9.4 \\
& 7\% DNS & $3.12~\te{-3}$ & $9.66~\te{-5}$ &  $1.33~\te{5}$ & 12.8 & 128 & 77.8 & 5.8\\
& 7\% LES & $3.27~\te{-3}$ & $8.99~\te{-5}$ &  $1.27~\te{5}$ & 11.4 & 128 & 73.5 & 5.5\\
& 14\% DNS &$2.99~\te{-3}$ & $1.03~\te{-4}$ &  $1.39~\te{5}$ & 14.3 & 127 & 59.5 & 3.3\\
& 14\% LES &$3.11~\te{-3}$ & $1.01~\te{-4}$ &  $1.33~\te{5}$ & 13.5 & 128 & 55.6 & 3.1\\
& 21\% DNS &$3.12~\te{-3}$ & $1.11~\te{-4}$ &  $1.33~\te{5}$ & 14.8 & 126 & 44.4 & 1.9\\
& 21\% LES &$3.20~\te{-3}$ & $1.06~\te{-4}$ & $1.30~\te{5}$ & 13.7 & 125 & 42.3 & 1.8 \\
& 29\% LES & $3.08~\te{-3}$ & $1.02~\te{-4}$ & $1.35~\te{5}$ & 13.7 & 129 & 31.4 & 1.0\\
& 36\% LES & $3.12~\te{-3}$  & $9.42~\te{-5}$ & $1.33~\te{5}$ & 12.5 & 125 & 23.7 & 0.6 \\
& 43\% LES & $3.16~\te{-3}$  & $9.00~\te{-5}$ & $1.31~\te{5}$ & 11.8 & 119 & 17.7 & 0.3 \\
Midpath & 50\% LES & $3.06~\te{-3}$  & $8.72~\te{-5}$ & $1.36~\te{5}$ & 11.8 & 125 & 14.3 & 0.2\\
Earth  & & & & $\mathcal{O}(\te{5} \ut{yr}$) & $\approx 10$ & $\approx 130$ & $\approx 2$ & $1/365$\\
\hline
\end{tabular}
\vspace*{0.1cm}\setstretch{1}\normalsize
\caption{\label{SVSA}Master time scales for secular variation and acceleration, together with dimensional time scales obtained after adjusting $\tau_{\mathrm{SV}}^{1}$ to the value 415 yr estimated in \cite{Lhuillier2011b}. Estimates for Earth's core are obtained from \cite{Gillet2010,Christensen2012} and A17.}
\end{center}
\end{table}

\begin{figure}
\centerline{\includegraphics[width=8cm]{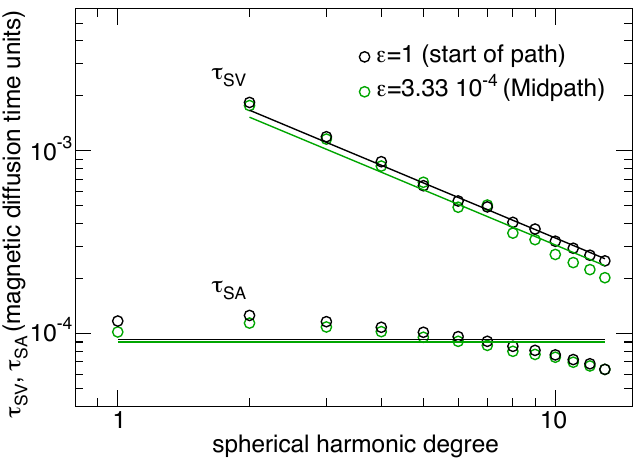}}
\caption{\label{tauSVSA}Secular variation and acceleration time scales $\tau_{\mathrm{SV}}(\ell)$, $\tau_{\mathrm{SA}}(\ell)$ as functions of the spherical harmonic degree $\ell$, in the \rev{start-of-path} and Midpath model. Solid lines represent the best fits of the form $\tau_{\mathrm{SV}}(\ell)=\tau_{\mathrm{SV}}^{1}/\ell$ and $\tau_{\mathrm{SA}}(\ell)=\tau_{\mathrm{SA}}^{0}$, with the master time scales $\tau_{\mathrm{SV}}^{1}$, $\tau_{\mathrm{SA}}^{0}$ reported in Table \ref{SVSA}.}
\end{figure}

At any given \rev{instant} in time $t$, spectra $E_{B}(\ell,t)$, $E_{\mathrm{SV}}(\ell,t)$ and $E_{\mathrm{SA}}(\ell,t)$ can be constructed that indicate the contribution from each spherical harmonic degree $\ell$ to the magnetic field energy $\vecB^{2}$, secular variation energy $(\partial\vecB/\partial t)^{2}$ and secular acceleration energy $(\partial^{2}\vecB/\partial t^{2})^{2}$ at the core-mantle boundary. Altitude-independent, scale-dependent secular variation \citep{Hulot1994,Lhuillier2011b} and acceleration \citep{Holme2011,Christensen2012} time scales can then be constructed as
\begin{eqnarray}
\tau_{\mathrm{SV}} (\ell) &=& \displaystyle \sqrt{\left[ E_{\mathrm{B}}(\ell,t)\right]/ \left[  E_{\mathrm{SV}}(\ell,t) \right]},\\
\tau_{\mathrm{SA}} (\ell) &=& \displaystyle\sqrt{\left[ E_{\mathrm{SV}}(\ell,t)\right]/ \left[ E_{\mathrm{SA}}(\ell,t)\right]},
\end{eqnarray}
where the square brackets denote an average in time. The function $\tau_{\mathrm{SV}}(\ell)$ has been shown to present a common structure in both geomagnetic field models and numerical dynamo simulations \citep{Lhuillier2011b}, that is parameterised using the form $\tau_{\mathrm{SV}}(\ell)=\tau_{\mathrm{SV}}^{1}/\ell$. Though the determination of $\tau_{\mathrm{SA}}(\ell)$ is intrinsically more uncertain, it has been proposed \citep{Christensen2012} that at large scales (typically $\ell\le 10$) $\tau_{\mathrm{SA}}(\ell)$ approaches a constant $\tau_{\mathrm{SA}}^{0}$, while at smaller scales $\tau_{\mathrm{SA}}(\ell)$ decreases like $1/\ell$. Throughout the path, the results confirm these two trends (Fig. \ref{tauSVSA}), though it should be noted that concerning the $1/\ell$ trend for $\tau_{\mathrm{SV}}(\ell)$ the Midpath model shows more significant deviations than the \rev{start-of-path} model. Here $\tau_{\mathrm{SV}}^{1}$ is determined by fitting this last trend to $\tau_{\mathrm{SV}}(\ell)$ between $\ell=2$ and $\ell=13$, and $\tau_{\mathrm{SA}}^{0}$ is determined by averaging $\tau_{\mathrm{SA}}(\ell)$ between degrees $\ell=1$ and $\ell=13$. Table \ref{SVSA} shows that the values of $\tau_{\mathrm{SV}}^{1}$ and $\tau_{\mathrm{SA}}^{0}$ normalised by the magnetic diffusion time $\tau_{\eta}$ remain approximately constant throughout the path.

It has become common practice to use non-canonical units to recast the dimensionless numerical model output back to the dimensional world \citep[e.g.][]{Aubert2015}. The goal is to provide useful comparisons with observations, despite the distance in parameter space between numerically feasible models and Earth's core conditions. The path theory gives support to this approach because the end of path matches the Earth's core estimates of all relevant quantities. Here time is rescaled by adjusting $\tau_{\mathrm{SV}}^{1}$ to Earth's core estimate 415 yr provided by \cite{Lhuillier2011b}. The resulting dimensional key time scales are reported in Table \ref{SVSA}. This rescaling is essentially equivalent to a diffusive scaling where the magnetic diffusion time scale is adjusted to a value $\tau_{\eta}\approx 1.3~\te{5} \ut{yr}$, because $\tau_{\mathrm{SV}}^{1}/\tau_{\eta}$ is about constant throughout the path. This latter value of $\tau_{\eta}$ corresponds to a magnetic diffusivity $\eta\approx 1.2 \ut{m^{2}/s}$ which stands at the mid-point of current estimates (see A17). Once time is rescaled, the overturn time $\tau_{U}\approx 130 \ut{yr}$ is about constant, at a value of about one third of $\tau_{\mathrm{SV}}^{1}= 415 \ut{yr}$, a characteristic property of models respecting $Rm\approx 1000$ \citep{Aubert2011,Bouligand2016}. The SA time scale $\tau_{\mathrm{SA}}^{0}$ is also constant to within $\pm 10\%$ at a value of about $\tau_{\mathrm{SA}}^{0}\approx 10 \ut{yr}$ that matches Earth's core estimates, a striking preliminary result given the expected change in dynamics along the path. This will be analysed in more depth in section \ref{struct}. Note that adjusting the SV time scale does not necessarily set the SA time scale, as these are two independantly evolving quantities \citep{Christensen2012}. The Alfvén time scale $\tau_{A}$ decreases and becomes increasingly separated from $\tau_{U}$ along the path, to become decadal at Midpath conditions. Finally, the planetary rotation period $2\pi\tau_{\Omega}$ decreases along the path towards its value in Earth's core, and it should be noted that rotational dominance on interannual time scales is not enforced at the start of path where $2\pi\tau_{\Omega}\approx 10 \ut{yr}$, and is only realised after 30\% of the path where $2\pi\tau_{\Omega}\le 1 \ut{yr}$. 

Length is rescaled to the dimensional world by using the canonical value $D=2260\ut{km}$. The time and length rescalings then provide the rescaling rules for the velocity field $\vecu$, and also for the Alfvén velocity $\vecB/\sqrt{\rho\mu}$ used in Figs. \ref{torsionalwaves},\ref{hovdvp},\ref{wavebal}. This in principle also naturally provides a rescaling for the magnetic field $\vecB$, but in this case the field amplitude is then Earth-like only at the end of the path where the Alfvén number $A$ is correct. To facilitate comparison of the model output with geomagnetic observations, in Figs. \ref{pathevol}-\ref{pulseSA} the magnetic field, SV and SA amplitudes are rather presented by setting the root-mean squared amplitude $B$ of $\vecB$ to its estimate $B=4\ut{mT}$ in Earth's core \citep{Gillet2010}. Since the field amplitude expressed in Elsasser units is approximately constant throughout the path i.e. $B/\sqrt{\rho\mu\Omega\eta} \approx 4.5$ (A17), this is equivalent to an Elsasser rescaling where the unit $\sqrt{\rho\mu\Omega\eta}$ is set to the value 0.9 mT. A similar reasoning applies to the rescaling of the density anomaly field $C$, where the chosen time and length units cannot be used except at the end of the path where the Rayleigh number $Ra_{F}$ is correct. Density anomaly is therefore rescaled by using the above rescaling rule for velocity and adjusting the dimensionless convective power in the shell to the value $P=3 \ut{TW}$ corresponding to the end of the path and to an estimate for Earth's core (see A17). 

\subsection{\label{numerical}Numerical Implementation details.}

The numerical implementation is described in A17, where all the relevant details can be found. A decomposition of $\vecu$, $\vecB$, $C$ in spherical harmonics up to degree and order $l_\mathrm{max}$ is used, together with a discretisation in the radial direction on a second-order finite-differencing scheme with a number of radial grid points $NR$. The grid parameters $l_\mathrm{max}$ and $NR$ have been reported in A17, at the exception of the direct (fully resolved) numerical simulation at 21\% of the path, which is original to this study and where $NR=720$, and $l_\mathrm{max}=320$. The spherical harmonics transform library SHTns is used \citep[][freely available at {\tt https://bitbucket.org/nschaeff/shtns}]{Schaeffer2013}. Time stepping is of second-order, semi-implicit type. In addition to fully resolved direct numerical simulations (type DNS in Tables \ref{tsratios},\ref{SVSA}), large-eddy simulations are used at advanced path positions (LES), by applying an hyperdiffusive treatment (see A17 for details) on $\vecu$ and $C$ in order to obtain numerically tractable models that maintain the same value $l_\mathrm{max}=133$ as at the \rev{start of path}. The physical relevance and accuracy of this treatment have been demonstrated in A17. In particular, it has been shown that this approximation does only weakly alter the large-scale structure of the fields $\vecu$, $\vecB$, $C$ and the large-scale force balance structure (compare the \rev{start-of-path} DNS and Midpath LES in Fig. \ref{forcebal}). All models produce a self-sustained magnetic field with a dominant axial dipole and without polarity reversals (see A17). The magnetic field morphology is also Earth-like throughout the explored part of the path. \rev{This is} attested by values $\chi^{2}\le 2.1$ of the diagnostic quantity introduced in \cite{Christensen2010}. \rev{This quantity is constructed from several criteria comparing the numerical dynamo output to the geomagnetic field, with values less than 2 characterising a good level of morphological resemblance.}

The determination of SV and SA requires long, high-cadence records of the magnetic field, with a carefully determined temporal sampling rate. The equilibrated runs from A17 have been continued for the equivalent of $\Delta t=20000 \ut{yr}$ or 154 outer core overturns (except for the 21\% path DNS simulation where $\Delta t=1200 \ut{yr}$), and the magnetic field at the outer boundary has been sampled every 50 numerical time steps up to spherical harmonic degree and order 13. This corresponds to a variable spacing between samples, from 0.5 yr at the \rev{start of path} down to 0.02 yr for the Midpath model, in any case at least an order of magnitude shorter than the planetary rotation period $2\pi\tau_{\Omega}$ (Table \ref{SVSA}). Convergence of the SV and SA energies has been checked by doubling the sampling rate on short portions of the runs. 

\section{\label{results}Results.}

\subsection{\label{struct}Spatio-temporal structure of secular acceleration at the core-mantle boundary.}
\begin{figure}
\centerline{\includegraphics[width=17cm]{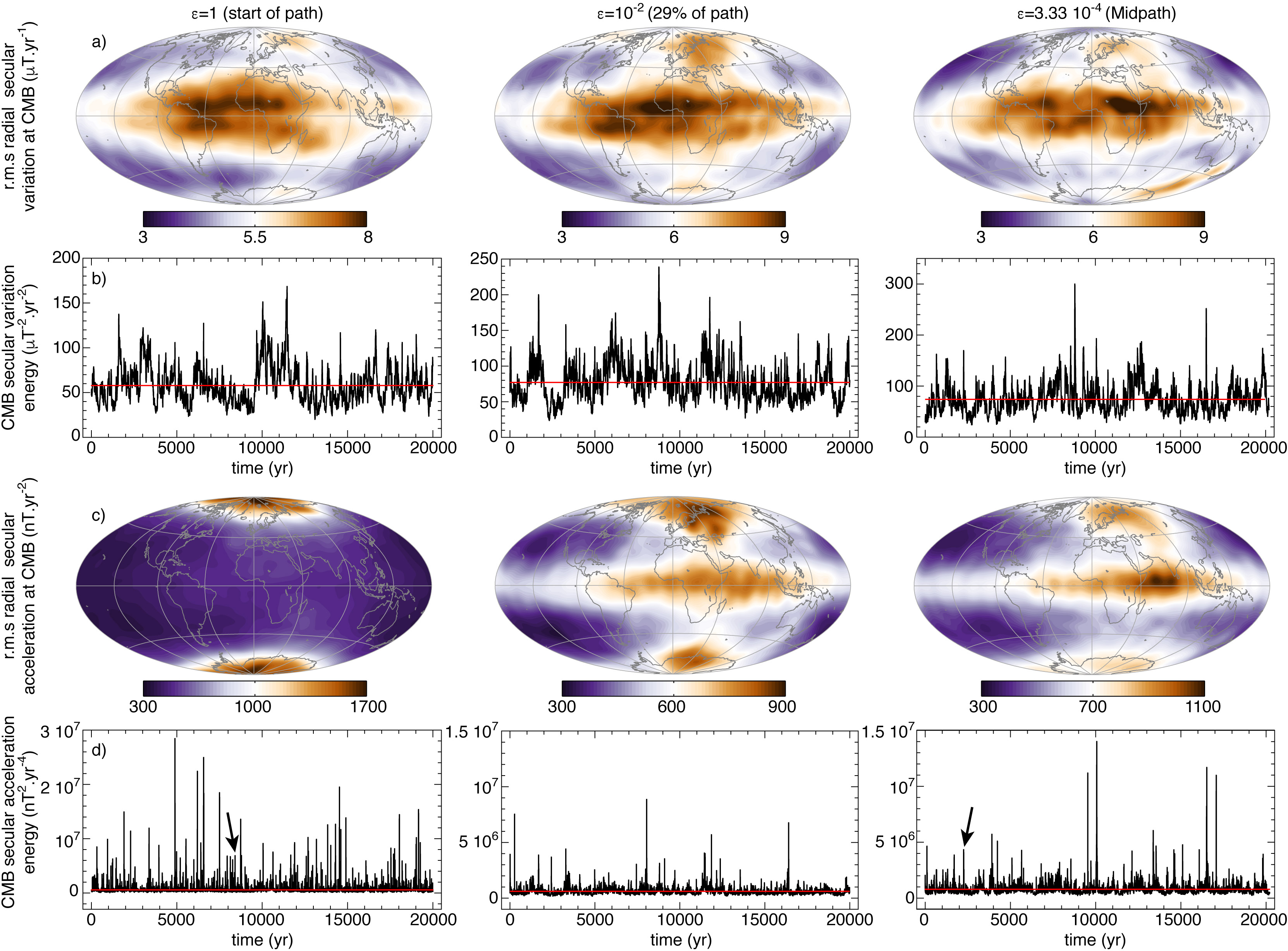}}
\caption{\label{pathevol} Evolution of SV and SA properties between three positions along the parameter space path: start (left), 29\%  (middle), and Midpath (right). Hammer projections of the r.m.s. amplitude of the SV (a) and SA (c) up to spherical harmonic degree 13 at the core-mantle boundary (CMB), together with corresponding time series of SV and SA energies (b,d). Red lines in (b,d) mark the time-averaged values of SV and SA energies. Arrows in (d) mark the two SA pulses analysed in detail in sections \ref{breakdown} and \ref{waves}.}
\end{figure}

Fig. \ref{pathevol} illustrates the properties and evolution along the parameter space path of the core-mantle boundary SV and SA truncated at spherical harmonic degree 13. SV patterns (Fig. \ref{pathevol}a) and average energies (Fig. \ref{pathevol}b) are approximately invariant throughout the path. This confirms the results obtained in \cite{Aubert2013b} concerning the localisation of SV at low latitudes and in an longitudinal hemispheric band within the Atlantic hemisphere, as a consequence of the bottom-up heterogeneous driving of convection from the inner core. This also further strengthens the case, initially made in  A17, for invariance of the dynamo kinematics along the path.

\begin{figure}
\centerline{\includegraphics[width=8cm]{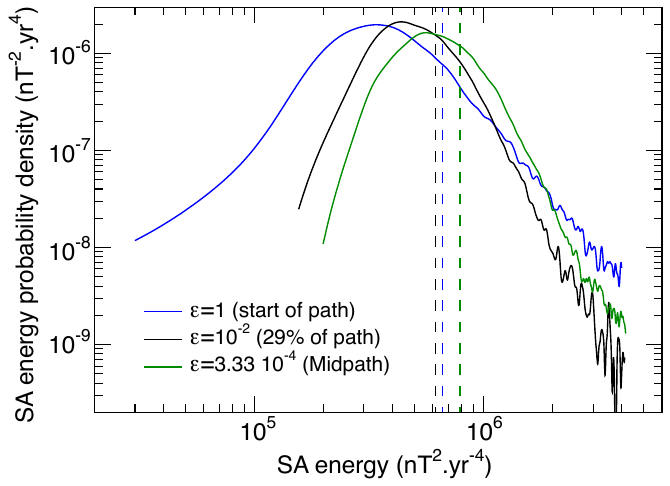}}
\caption{\label{hists} \rev{Probability density functions describing the statistical distribution of SA energy for the start-of-path, 29\% of path and Midpath models, as obtained from the SA time series of Fig. \ref{pathevol}d. The average SA energy for each model is represented as a dashed vertical line.}}
\end{figure}

The SA average energy is also approximately invariant along the path \rev{(Figs. \ref{pathevol}d, \ref{hists})}, as anticipated from the invariant value of $\tau_{\mathrm{SA}}$ in Table \ref{SVSA}. However, significant changes are observed in the spatial localisation of SA (Fig. \ref{pathevol}c). At \rev{the start of path,} the SA pattern is dominated by activity at the poles, with no apparent influence of longitudinally heterogeneous convection. This gradually evolves towards a stable pattern observed from 30\% of the path onwards where dominant SA activity is found both in high-latitude regions and in an equatorial band, with clear longitudinal heterogeneities inherited from the convective pattern. Equatorial dominance in the SA pattern is observed in the Midpath model, while polar dominance remains at 29\% of the path, though either of these should be relativised given the weak relative dominance (the difference in amplitude between pole and equator being about 15\% in both cases) and the limited time span of the high-cadence records. 

\rev{Along the path, significant changes in the temporal variability of the SA energy can also be observed in Fig. \ref{pathevol}d and quantified through probability density functions (Fig. \ref{hists}). At the start of path, the SA energy distribution is extremely broad (Fig. \ref{hists}) as it spans three orders of magnitude (the highest SA energies are not represented in Fig. \ref{hists} because of the rarity of events). In the SA energy range below the average value, the distribution becomes considerably narrower at 29\% of the path, and only weakly evolves afterwards. In the SA energy range above the average value,} intermittent short pulses of energy up to 40 times the average can be  observed throughout the path (Fig. \ref{pathevol}d), with a typical duration on the order of a few years in all cases. The pulses are strongest and most frequent at the start of path \rev{(Figs. \ref{pathevol}d,\ref{hists})}, and their strength and frequency gradually decreases to reach a minimum at 29\% of the path, after which they gradually increase again. Due to their intermittency, pulses do not influence the average SA energy (\rev{Fig. \ref{hists}}) nor the r.m.s. SA localisation patterns (Fig. \ref{pathevol}c). Their spatial localisation however tends to respect these patterns. At \rev{the start of path,} all pulses of energy larger than $4~\te{6} \ut{nT^{2}.yr^{-4}}$ in the sequence were found near the poles. At 29\% of the path events of energy larger than $3~\te{6} \ut{nT^{2}.yr^{-4}}$ were equipartitioned between high ($45^{\circ}$ and higher) and low latitudes. In the Midpath model sequence 80\% of the events of energy larger than $4~\te{6} \ut{nT^{2}.yr^{-4}}$ were found at low latitudes, and 60\% of the events were found at an elevation from equator less than $20^{\circ}$. These first results \rev{suggest that important dynamical changes take place in the system along the path (especially between the start and 29\%), but they also show that} an integral measure such as $\tau_{\mathrm{SA}}$ is essentially insensitive to these  dynamical changes.

\begin{figure}
\centerline{\includegraphics[width=14cm]{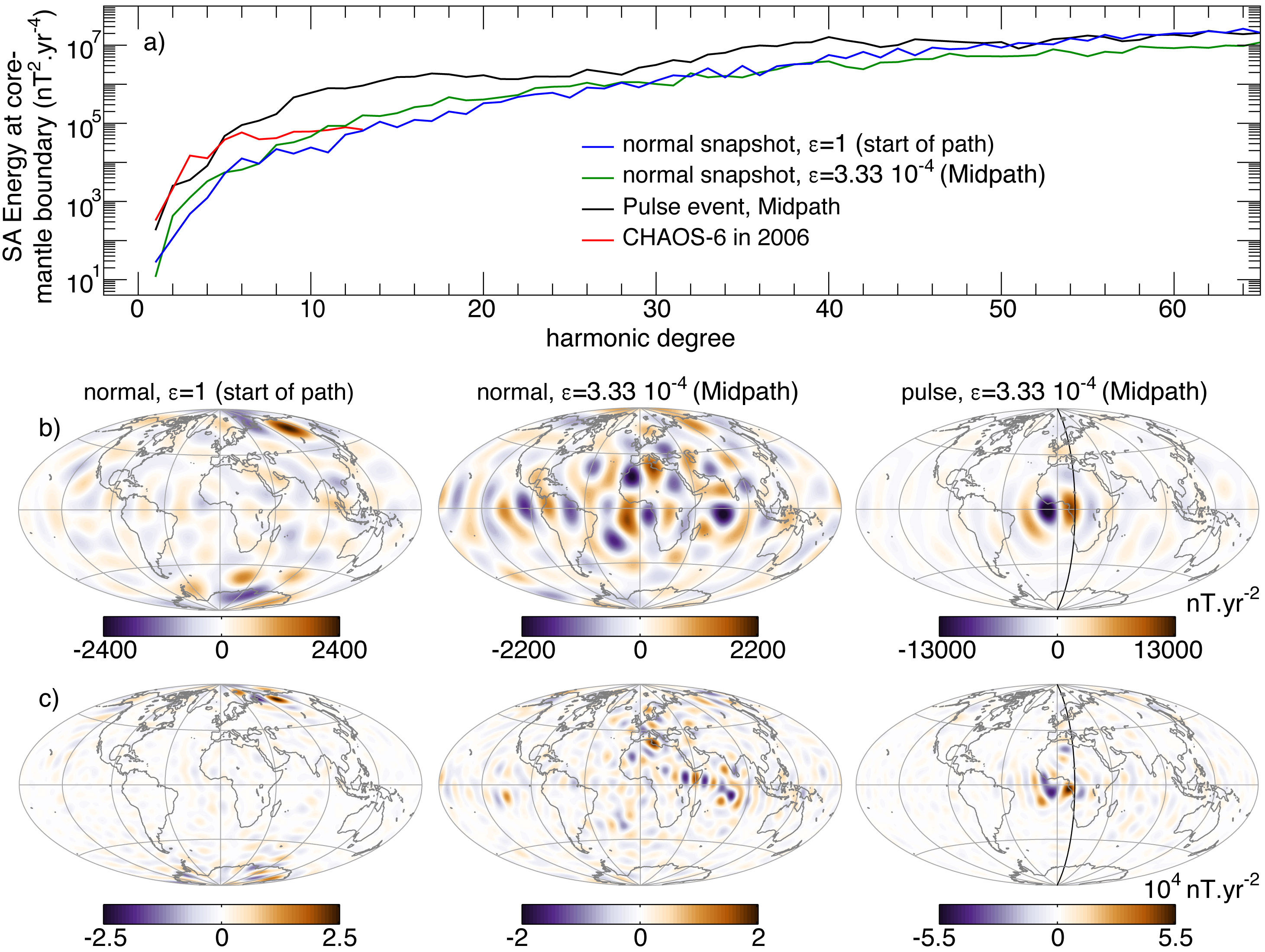}}
\caption{\label{hires} a: Core-mantle boundary spatial SA energy spectra for typical snapshots of the \rev{start-of-path} and Midpath models, and for the pulse event located by an arrow in Fig. \ref{pathevol}d for the Midpath model. b,c: Hammer projections of the radial SA at the core-mantle boundary (orange is outwards) for the same events, truncated at spherical harmonic degree 13 (b) and \rev{30} (c). The black curves delineate the longitude selected for analysis of the Midpath pulse event in section \ref{waves}.}
\end{figure}

The high-cadence records that served to produce Fig. \ref{pathevol} are truncated at spherical harmonic degree 13, already beyond the resolution of modern geomagnetic field models \citep[e.g.][]{Finlay2016b}, but below that used in the numerical dynamo models. It is instructive to examine in Fig. \ref{hires} the morphology of core-mantle boundary SA snapshots at various truncation levels. The SA energy spatial spectrum (Fig. \ref{hires}a) is blue i.e. dominated by energy at small-scales in all cases, as found previously in numerical simulations \citep{Christensen2012}, and in geomagnetic field models such as CHAOS-6 \citep{Finlay2016b}. At large scales, the spectrum of the pulse event in the Midpath model also quantitatively matches that obtained in CHAOS-6 at the epoch 2006 where a SA pulse has been identified \citep{Chulliat2010}. Comparing the two spectra also suggests a deficit in power beyond degree 8 in CHAOS-6, presumably related to the limit of this field model as regards the resolution of SA. The blue spectrum is problematic in the sense that one cannot be certain to obtain a reliable low-resolution image of the SA, since the part of energy that is discarded through truncation always dominates the part that is retained. Considering a similar problem for the SV, \cite{Holme2011} conjectured that the signal could nevertheless present a good degree of coherence i.e. the localisation of features observed at large scales could be preserved in higher-resolution images. Our results for SA broadly confirm this view (Fig. \ref{hires}b,c), with the higher-resolution images (here truncated at degree \rev{30, with similar results also obtained when truncating at degree 60}) generally preserving the localisations observed in the low-resolution versions (truncated at degree 13), though some small-scale features are inevitably misrepresented. The coherence between low- and high-resolution maps is especially clear during the SA energy pulse event of the Midpath model (compare the right column in Fig. \ref{hires}b,c). 

\begin{figure}
\centerline{\includegraphics[width=8cm]{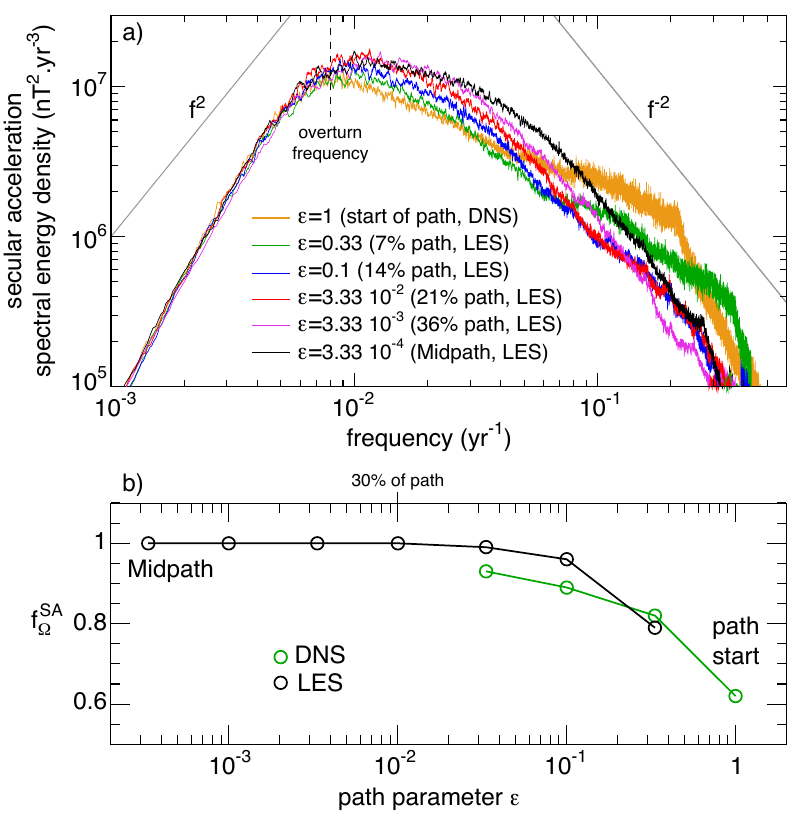}}
\caption{\label{psdrotdom} a: Spectral density in the frequency domain of the SA energy at the core-mantle boundary for models taken along the parameter space path. The dashed vertical line marks the overturn frequency $f_{U}=1/\tau_{U}$ which is constant along the path (Table \ref{SVSA}). The grey slanted lines mark the spectral indices $\pm 2$. b: Fraction $\mathrm{f}_{\Omega}^\mathrm{SA}$ of SA energy contained within the rotationally-dominated frequency range $[0,\Omega/2\pi]$ as a function of the path parameter $\epsilon$ (Earth's core conditions are towards the left of this graph).}
\end{figure}

We have seen that the SA energy at the core-mantle boundary is approximately preserved along the path (Fig. \ref{pathevol}) but presents variable spatial heterogeneities. It is also interesting to examine the variability in the temporal frequency domain (Fig. \ref{psdrotdom}a). \revm{To this end, the core-mantle boundary SA energy $E_{\mathrm{SA}}$ is decomposed by using a Thomson multitaper method with concentration half-bandwidth $W=4/\Delta t=2~\te{-4} \ut{yr^{-1}}$ \citep[see][]{Bouligand2016}, such that the overturn frequency $f_{U}\approx 1/130 \ut{yr^{-1}}$ (table \ref{SVSA}) and higher frequencies are well outside the unresolved range $[0,W]$. This tool is useful to separate slow and rapid dynamics (respectively $f<f_{U}$ and $f>f_{U}$) in the frequency domain. This also enables (Fig. \ref{psdrotdom}b) the evaluation of the fraction $\mathrm{f}_{\mathrm{SA}}^{\Omega}$ of SA power contained within the rotationally-dominated frequency range (Fig. \ref{psdrotdom}b), i.e. at frequencies lower than the planetary rotation \rev{frequency} $f_{\Omega}=\Omega/2\pi$:
\begin{equation}
\mathrm{f}_{\mathrm{SA}}^{\Omega} = \dfrac{\int_{0}^{f_{\Omega}} E_{\mathrm{SA}}(f)\,\mathrm{d}f}{E_{\mathrm{SA}}}.
\end{equation}}
Fig \ref{psdrotdom}a shows that the SA spectral energy density is generally invariant at frequencies lower than that of convective overturn $f_{U}=1/\tau_{U}$, a confirmation of the invariance of slow convective dynamics along the path (see A17). Similar to the trends observed in Fig. \ref{pathevol}, considerable variability is however observed between the \rev{start of path} and $30\%$ of the path ($\epsilon\ge \te{-2}$, Fig. \ref{psdrotdom}a), at frequencies ranging from $f_{U}$ to $4~\te{-1} \ut{yr^{-1}}$. There the high-frequency decay of the SA spectral energy density curve becomes significantly steeper and gradually approaches a $f^{-2}$ behaviour. The most important dynamical change taking place in this frequency range is the enforcement of rotational dominance as we progress along the path. Indeed, from \rev{the start of path} to 30\% of the path the planetary rotation frequency evolves from $f_\mathrm{\Omega}=\te{-1} \ut{yr^{-1}}$ to $f_\mathrm{\Omega}=1 \ut{yr^{-1}}$ (Table \ref{SVSA}), implying a larger rotationally-dominated range (i.e. frequencies such that $f\le f_\mathrm{\Omega}$) in the latter case. As a consequence, a significant amount of SA energy density is not rotationally-dominated at \rev{the start of path} (see values of $\mathrm{f}_{\Omega}^\mathrm{SA}$ significantly below 1 in Fig. \ref{psdrotdom}b), while the energy is entirely contained within the rotationally-dominated range in the rapidly rotating regime obtained at 30\% of the path. As we advance beyond this point, the slope of the spectral decay at high frequencies remains close to $f^{-2}$, while at intermediate frequencies $f_{U}<f<\te{-1} \ut{yr^{-1}}$ we observe the appearance and gradual extension of a flatter spectral energy density range. Note that a spectral decay less steep than $f^{-2}$ implies that the time derivative of SA has a $f^{\alpha}$ profile with $\alpha\ge 0$, another indication of the possible occurrence of rapid and intermittent energy pulses (Fig. \ref{pathevol}d).

\subsection{\label{breakdown}Contributions to the secular acceleration.}
\begin{figure}
\centerline{\includegraphics[width=12cm]{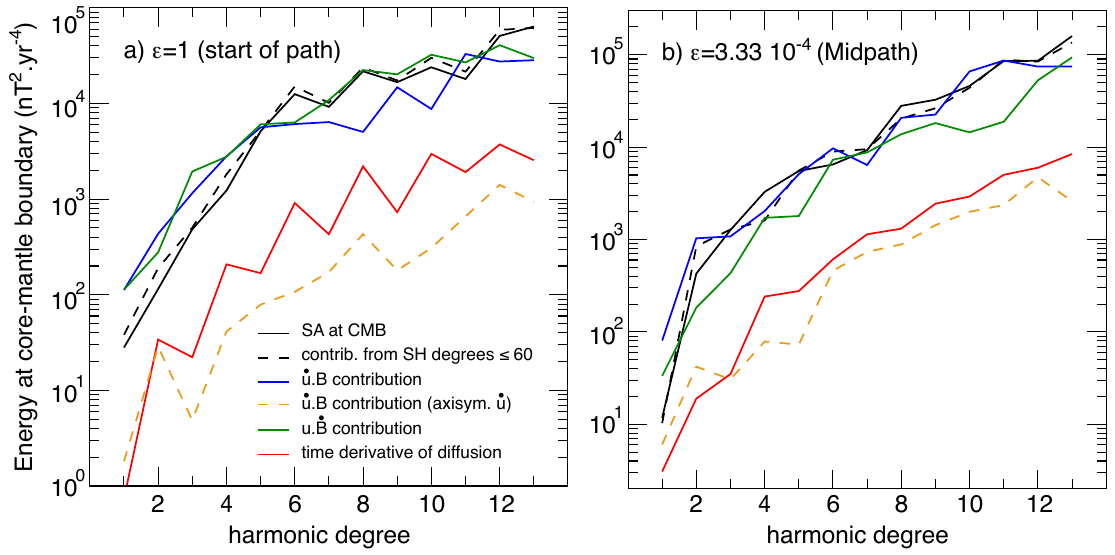}}
\caption{\label{normalSA}Core-mantle boundary spatial SA energy spectra (black) up to spherical harmonic degree 13 for typical snapshots of (a) the \rev{start-of-path} and (b) the Midpath models. Also presented are the contributions from (see equation \ref{MISA}): the interaction of $\partial\vecu/\partial t$ and $\vecB$ (blue), both fields considered up to degree 60; the interaction of $\vecu$ and $\partial\vecB/\partial t$ (green), both fields again considered up to degree 60; the time derivative of magnetic diffusion (red) up to degree 13. The dashed black line is the sum of these three contributions, and indicates the error made by not considering the contributions from field and flow beyond degree 60. The dashed orange line presents the contribution from axisymmetric flow acceleration (harmonic order 0 and again degree up to 60) to the SA.}
\end{figure}

\revm{The core-mantle boundary SA can be broken down according to the three main contributions, obtained by writing the time derivative of the magnetic induction equation (equation 2.2 in A17), here in dimensional form:
\begin{equation}
\dfrac{\partial^{2} \vecB}{{\partial t}^{2}}=\nabla\times\left(\ddp{\vecu}{t}\times \vecB\right)+\nabla\times\left(\vecu\times \ddp{\vecB}{t}\right)+\eta\boldsymbol{\Delta} \left(\ddp{\vecB}{t}\right).\label{MISA}
\end{equation}
From left to right, the terms on the right-hand-side of (\ref{MISA}) respectively represent the action of the flow acceleration $\partial\vecu/\partial t$ on the magnetic field, the action of the flow $\vecu$ on the SV, and the diffusive contribution to the SA. At selected times, a SA budget is obtained by computing in the spectral space the contributions to SA up to degree 13 from diffusion and from the interaction of magnetic field, SV, flow and flow acceleration up to degree 60.} In Fig. \ref{normalSA} we evaluate these contributions for the typical snapshots of the \rev{start-of-path} and Midpath models shown in Fig. \ref{hires}. In both cases the SA up to degree 13 is essentially captured by considering the various interactions of flow and magnetic field up to degree 60. Diffusion is found to be negligible in both cases, in line with the analysis made in \cite{Christensen2012}. The contributions from $\vecu$ and $\partial\vecu/\partial t$ are generally balanced in snapshots (with a slight dominance of the latter in the Midpath case), and the actual secular acceleration tends to result from some degree of mutual cancellation. It is worth noting that the axisymmetric flow acceleration does contribute only 2 to 4\% of the total SA energy. 

\begin{figure}
\centerline{\includegraphics[width=12cm]{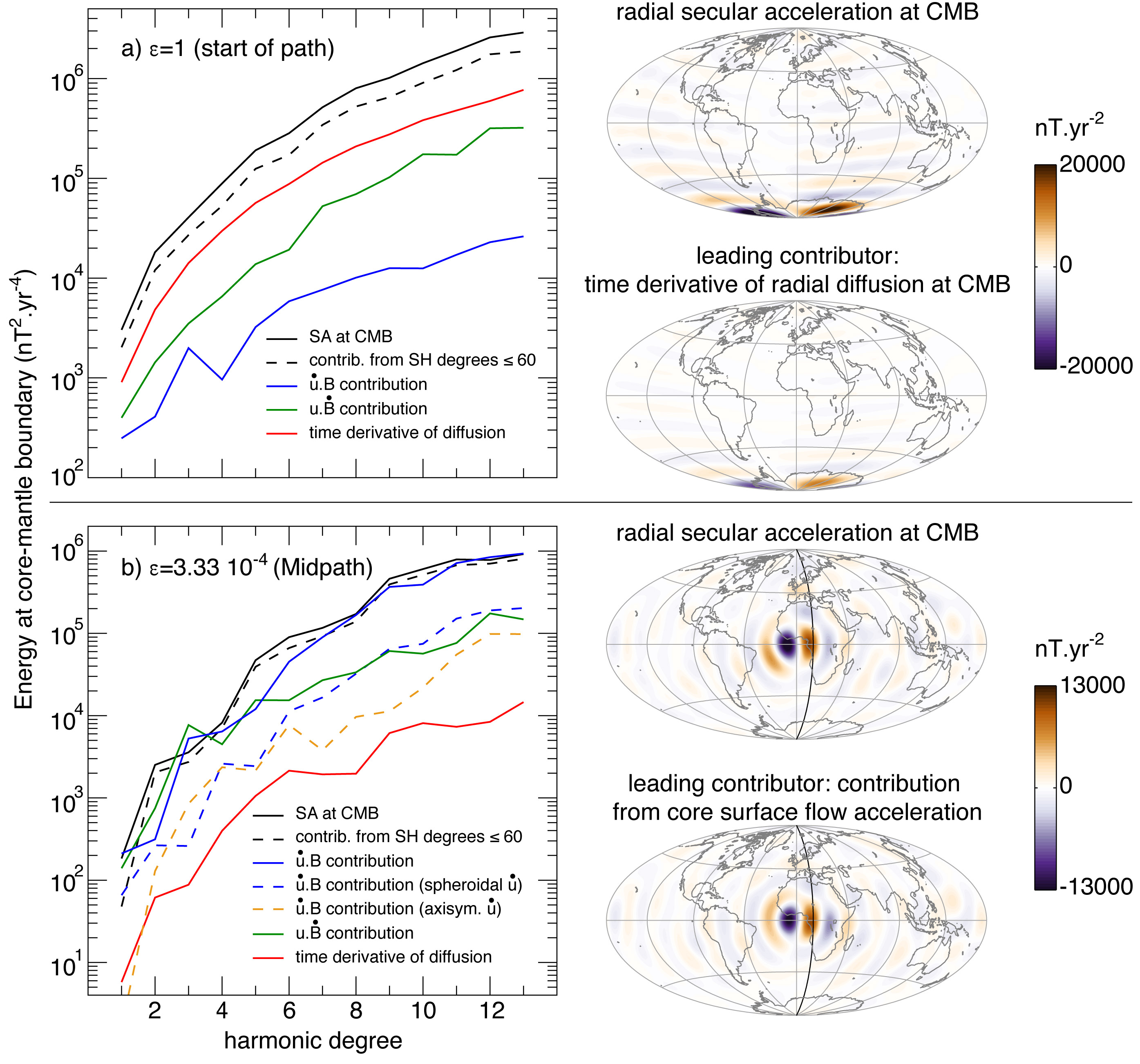}}
\caption{\label{pulseSA}Left: core-mantle boundary spatial SA energy spectra up to spherical harmonic degree 13, at pulse times marked by arrows in Fig. \ref{pathevol}d for the \rev{start-of-path} (a) and Midpath (b) models. See Fig. \ref{normalSA} for the definitions of the presented contributions. Here the contribution from the interaction of the spheroidal part of $\partial\vecu/\partial t$ with $\vecB$ is additionally reported (dashed blue line). Right: Hammer projections of the radial SA at the core-mantle boundary (orange is outwards), and the leading contribution to this SA. The black curves in (b) delineate the longitude selected for analysis of the pulse event in section \ref{waves}.}
\end{figure}

The situation changes significantly when considering the contributions to SA during pulse events (Fig. \ref{pulseSA}). Pulses at  \rev{the start of path systematically} present a polar structure (Fig. \ref{pulseSA}a) strongly deviating from a non-diffusive behaviour. The time derivative of magnetic diffusion is indeed now found as being the leading contributor, followed by the action of $\vecu$ on the secular variation. The effect of $\partial\vecu/\partial t$ is marginal. This indicates that the \rev{start-of-path} SA energy pulses can be explained as resulting from events of magnetic flux expulsion. In contrast, near-equatorial pulses at Midpath conditions (the most frequent configuration, Fig. \ref{pulseSA}b) remain non-diffusive, but \rev{systematically} show a marked dominance of the contribution of $\partial\vecu/\partial t$ over that from $\vecu$, particularly at the most energetic scales $\ell>5$. Within this contribution from $\partial\vecu/\partial t$, the parts arising from spheroidal (upwelling) flow acceleration and axisymmetric flow acceleration are both found to be subdominant (these parts respectively amount to 16\% and 8\% of the total SA energy). The equatorial pulses in the rapidly-rotating regime of the Midpath model are therefore primarily driven by non-axisymmetric, toroidal flow accelerations. Further examination also confirms that the azimuthal component of this toroidal flow acceleration dominates the latitudinal component, as can be expected from near-equatorial motion in a rapidly-rotating fluid.

Polar pulses in the Midpath model do not have a common origin with polar pulses observed at the \rev{start of path}, as they are found to be non-diffusive like their equatorial counterparts, hence discarding an explanation in terms of flux expulsion. However, our analysis beyond this is limited because they have been found to typically result from the interaction of field and flow beyond degree 60. This obscures the prospect of clearly identifying large-scale flow and magnetic field structures responsible for these less frequent events. 

\subsection{\label{waves}Rotationally-dominated hydromagnetic waves and their role in producing the secular acceleration.} 

\begin{figure}
\centerline{\includegraphics[width=8cm]{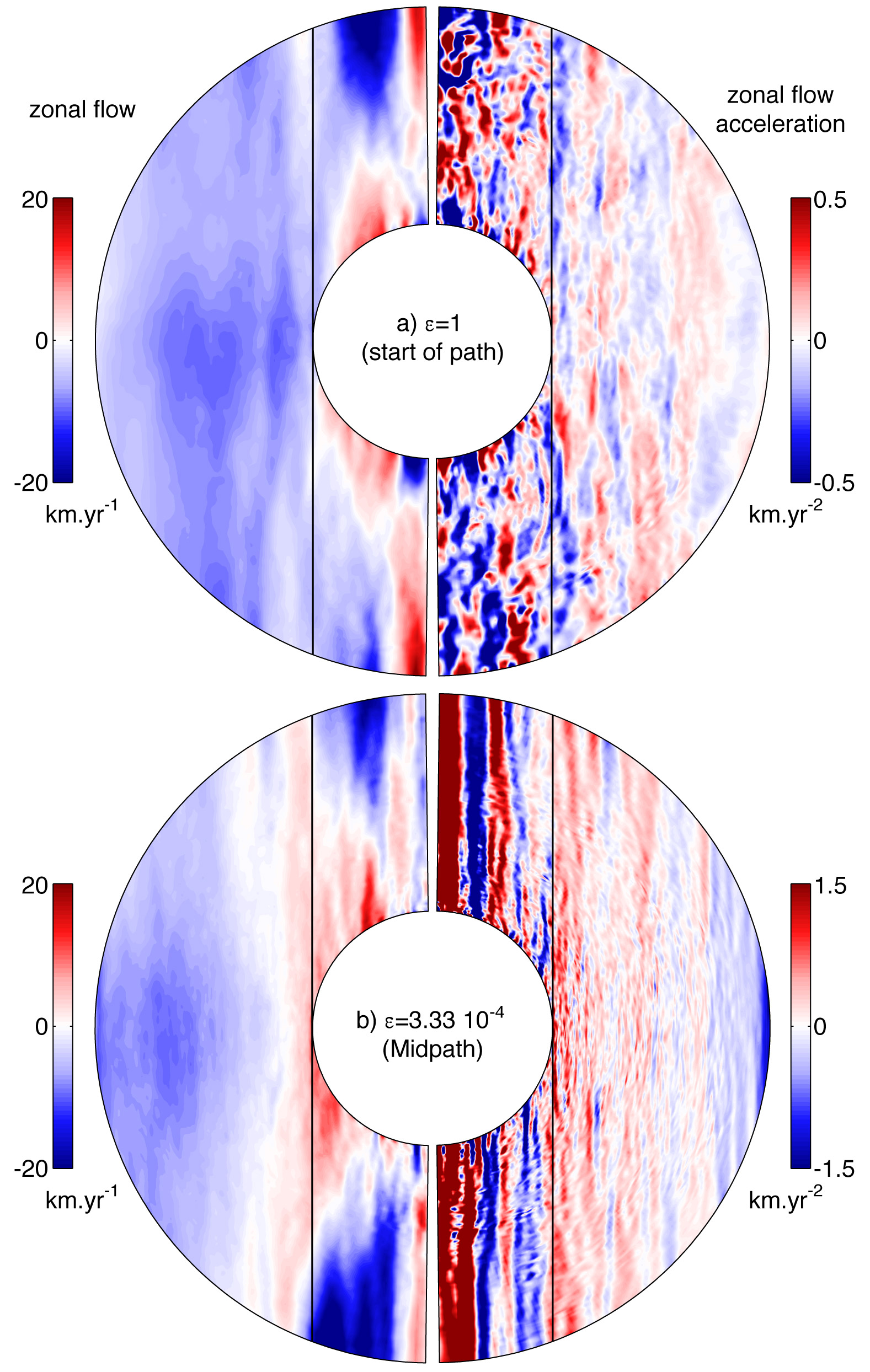}}
\caption{\label{2D3D}Meridional cuts of axisymmetric azimuthal (or zonal) flow snapshots (left, blue is westwards) and zonal flow acceleration snapshots (right), in the \rev{start-of-path} (a) and Midpath model (b). The vertical black lines mark the axial cylinder tangent to the inner core (the tangent cylinder).}
\end{figure}

\begin{figure}
\centerline{\includegraphics[width=11cm]{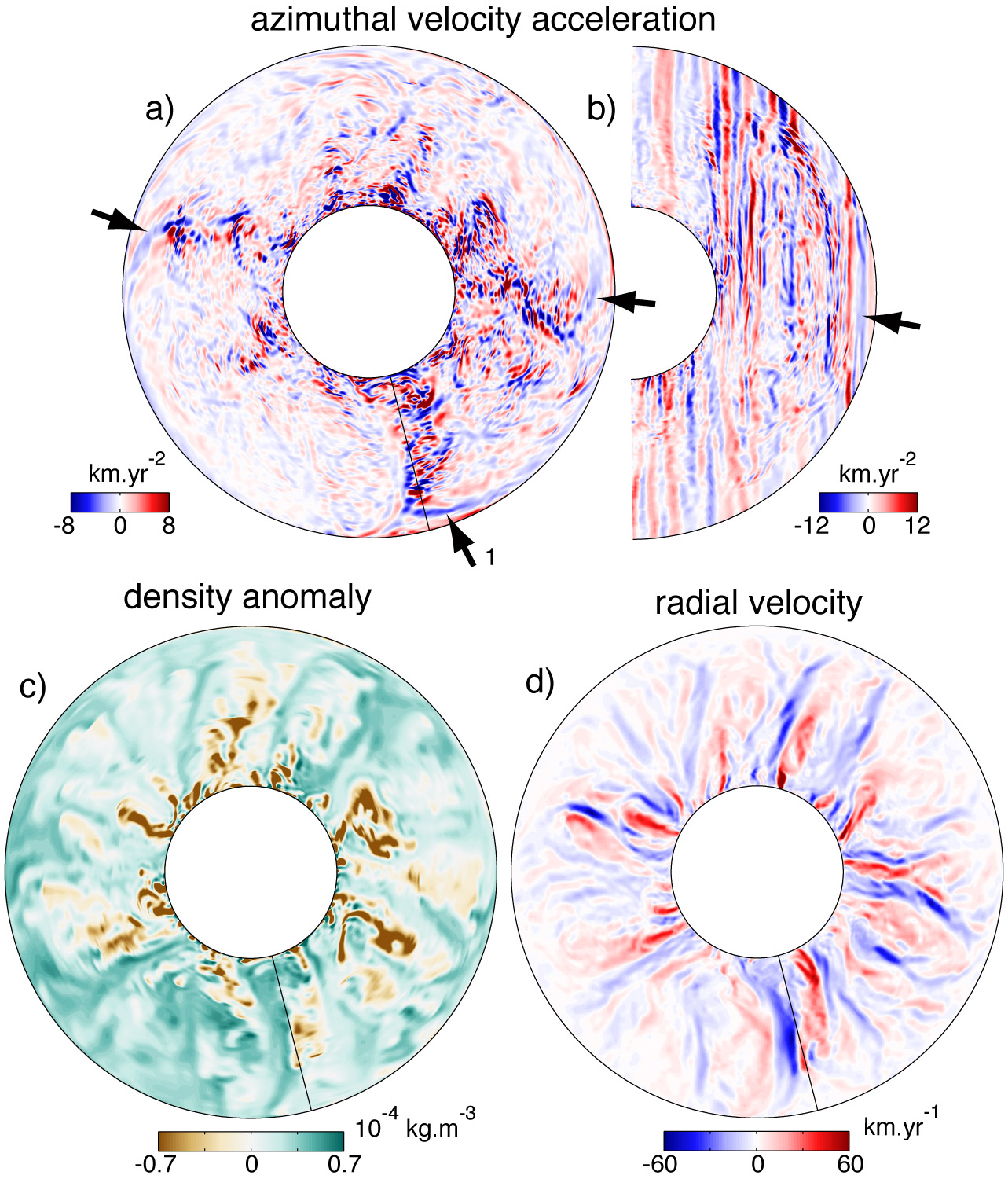}}
\caption{\label{planforms}Midpath model planforms obtained 8.7 years before the equatorial SA pulse located by an arrow in Fig. \ref{pathevol}d (right panel) and imaged in Fig. \ref{pulseSA}b. a: Equatorial planform of the azimuthal flow acceleration (blue is westwards). b: Meridional planform of azimuthal flow acceleration at the analysis longitude marked by a black line in (a,c,d). Arrows in (a,b) locate the quasi-geostrophic Alfv\'en (QGA) wave patterns (location 1 corresponds to the location of the SA pulse and is further examined in Fig. \ref{vorticities}). c: equatorial planform of density anomaly (orange denotes lighter fluid). d: equatorial planform of the radial velocity (red is outwards).}
\end{figure}

We next turn to the dynamical flow structures present in our models and their links with the secular acceleration patterns. Fig. \ref{psdrotdom} has presented evidence for the enforcement of the rotationally-dominated regime at path positions beyond 30\%. This can also be checked directly by examining the structure of flow acceleration in the physical space (Fig. \ref{2D3D}). \rev{Here the zonal (or axisymmetric azimuthal) flow $\overline{\vecu}\cdot\vc{e}_\varphi$ and its acceleration $\partial (\overline{\vecu}\cdot\vc{e}_\varphi)/\partial t$ are examined in the start-of-path and Midpath models ($\vc{e}_{\varphi}$ is the unit vector in the azimuthal direction, and the overbar represents an average taken in the azimuthal direction).} The zonal flow structure is preserved throughout the path (left panels), with a thermal wind pattern (A17) comprising westward motion outside the axial cylinder tangent to the inner core (the tangent cylinder) and polar vortices inside the tangent cylinder. This is another expression of the general kinematic invariance observed along the path. In contrast, the structure of zonal flow acceleration changes along the path, from a fully three-dimensional structure at \rev{the start of path} (Fig. \ref{2D3D}a, right panel) to a two-dimensional columnar structure in the Midpath model (Fig. \ref{2D3D}b). This axially columnar structure is the expression of the Proudman-Taylor constraint that follows from rotational dominance. The change to a two-dimensional structure also applies to non-axisymmetric azimuthal flow accelerations (see Fig. \ref{planforms}b below), though with patterns locally more perturbed by magnetic and buoyancy forces. In the Midpath model, significant columnar flow accelerations can also be observed at large cylindrical radii close to the equatorial core-mantle boundary (Fig. \ref{2D3D}b, \ref{planforms}a,b), \rev{while these accelerations are not present at the start of path (Fig. \ref{2D3D}a). The structural changes observed for flow acceleration along the path are consistent with the evolution of SA patterns obtained in Fig. \ref{pathevol}c. As we move along the path, the progressive loss of a spatial degree of freedom in flow acceleration reduces the efficiency of SA induction at high latitudes, while the appearance of flow acceleration at large cylindrical radii enhances the low-latitude SA.}

\begin{figure}
\centerline{\includegraphics[height=9cm]{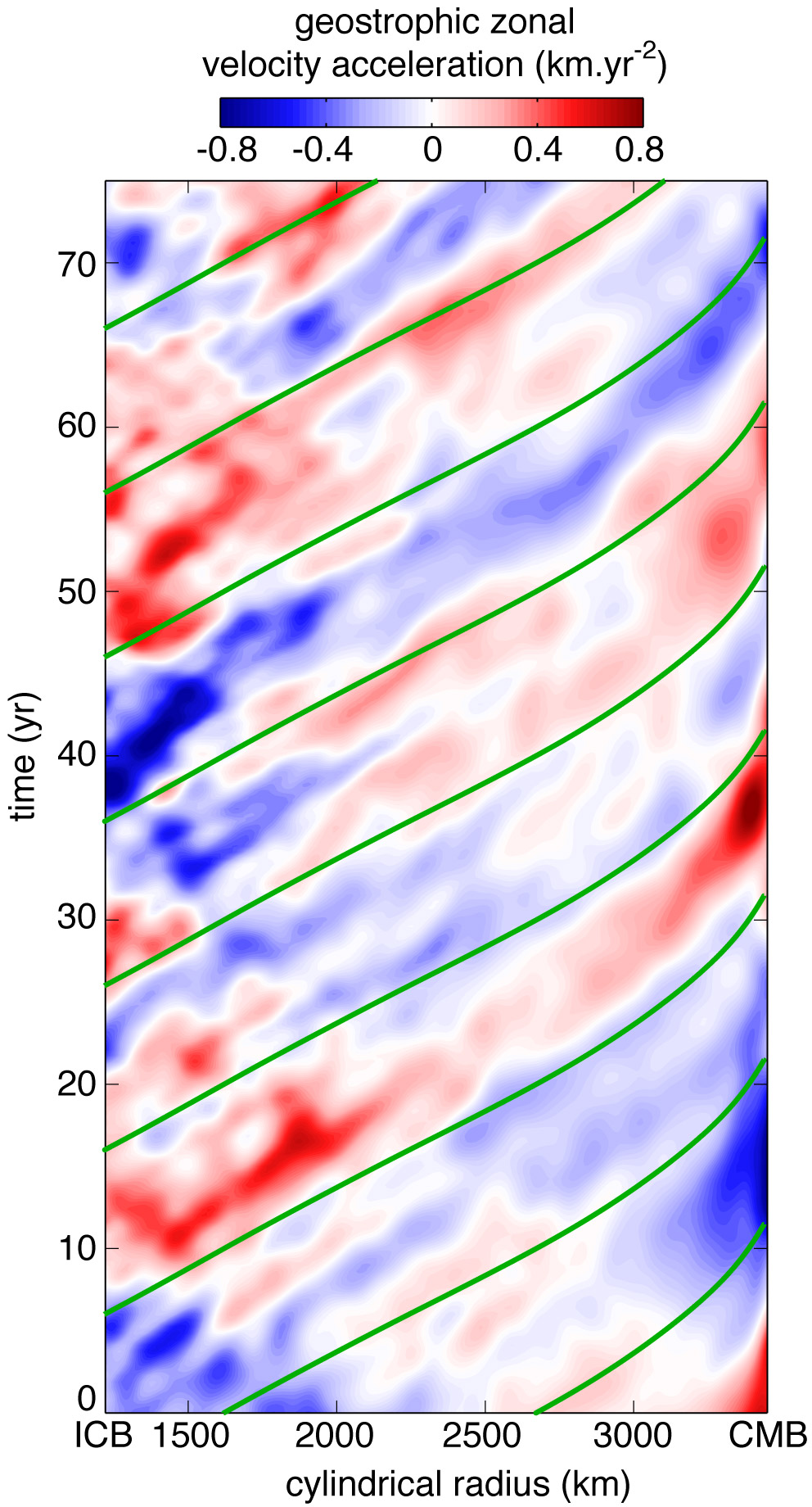}}
\caption{\label{torsionalwaves}Time-\rev{cylindrical} radius plot of the time derivative $\partial v_{g}/\partial t$ of the geostrophic velocity (equation \ref{geosvel}) outside the tangent cylinder in the Midpath model. Green lines represent the theoretical propagation tracks of torsional waves at the Alfv\'en speed $c_{A}(s,t)$ (equation \ref{TWvel}).}
\end{figure}

Along the path, the enforcement of rotational dominance is also accompanied with an increase in the amplitude of the MAC forces relative to inertia, as attested by Fig. \ref{forcebal} and the decrease in the Alfvén number $A$ (Table \ref{tsratios}) which is a proxy for the ratio of the inertial and magnetic forces (A17). From this follows an increase in forcing of magnetic origin, and the evolution of the Lundquist number $S$ presented in table \ref{tsratios} also signals a decrease of Alfvén wave attenuation along the path. Both factors imply that rotationally-dominated magneto-inertial waves should be more prominent at advanced path positions. An important class is geostrophic, columnar transverse torsional waves that magnetically couple axial cylinders in the shell and propagate in the cylindrical radial direction. Such waves cannot be exhibited at the conditions of the \rev{start of path} \citep[e.g.][]{Wicht2010}, as confirmed here from the non-columnar structure of zonal flow acceleration (Fig. \ref{2D3D}a). In contrast, in the Midpath model torsional waves are ubiquitous and can be exhibited in a straightforward manner by \rev{considering the geostophic zonal velocity $v_{g}$ defined as the average of the zonal flow along axial columns parallel to $\vc{\Omega}$:
\begin{equation}
v_{g} (s,t)= \overline{\left<\vecu\right>}\cdot\vc{e}_\varphi.\label{geosvel}
\end{equation}
Here the angled brackets denote the axial average defined as
\begin{equation}
\left< \vecu \right > (s,\varphi,t) =  \dfrac{1}{z_{+}-z_{-}}\int_{z_{-}}^{z_{+}} \vecu(s,\varphi,z,t)\, \mathrm{d} z,
\end{equation}
where $s,\varphi,z$ are cylindrical coordinates, and the vertical integral is evaluated between the lower and upper heights $z_{-,+}$ of an axial column at cylindrical radius $s$. Fig \ref{torsionalwaves} shows clear outwards propagation patterns in the geostrophic velocity acceleration $\partial v_{g}/\partial t$}. At all instants in space and time, the alternating wavefronts well match the theoretical \rev{Alfvén propagation velocity $c_{A}(s,t)$ constructed with the root-mean-squared amplitude of the magnetic field on axial cylinders \citep[e.g.][]{Schaeffer2017}
\begin{equation}
c_A (s,t) = \dfrac{1}{\sqrt{\rho\mu}} \left< \overline{(\vecB\cdot\vc{e}_{s})^{2}} \right>^{1/2},\label{TWvel}
\end{equation}
where $\vc{e}_{s}$ is the unit vector in the cylindrical radial direction.}
It is interesting to observe that in the regime reached by the Midpath model, waves and convection are clearly separated in the time domain in the sense that the propagation speed $c_{A}\approx 100 \ut{km/yr}$ significantly exceeds the typical \rev{unidirectional} velocity $D/\sqrt{3}\tau_{U}\approx 10 \ut{km/yr}$ of material advection, the ratio of the two typically representing the inverse of the Alfvén number $A$  (Table \ref{tsratios}). Similarly to \cite{Schaeffer2017} and also to observations made by \cite{Gillet2010}, the waves slow down as they approach the equatorial core-mantle boundary. The wave amplitude observed outside the tangent cylinder in the Midpath model is in agreement with that retrieved in Earth's core by \cite{Gillet2010}, typically 10 times weaker than that of non-axisymmetric flow accelerations inside the core (see Fig. \ref{planforms}a,b below). Torsional waves therefore cannot be expected to carry a dominant SA signature, as demonstrated by the contributions of axisymmetric flow acceleration in Figs. \ref{normalSA}b, \ref{pulseSA}b. This result gives a first rationalisation of the insensitivity of integral measures such as $\tau_{\mathrm{SA}}$ to the appearance of hydromagnetic waves. 

\begin{figure}
\centerline{\includegraphics[height=9cm]{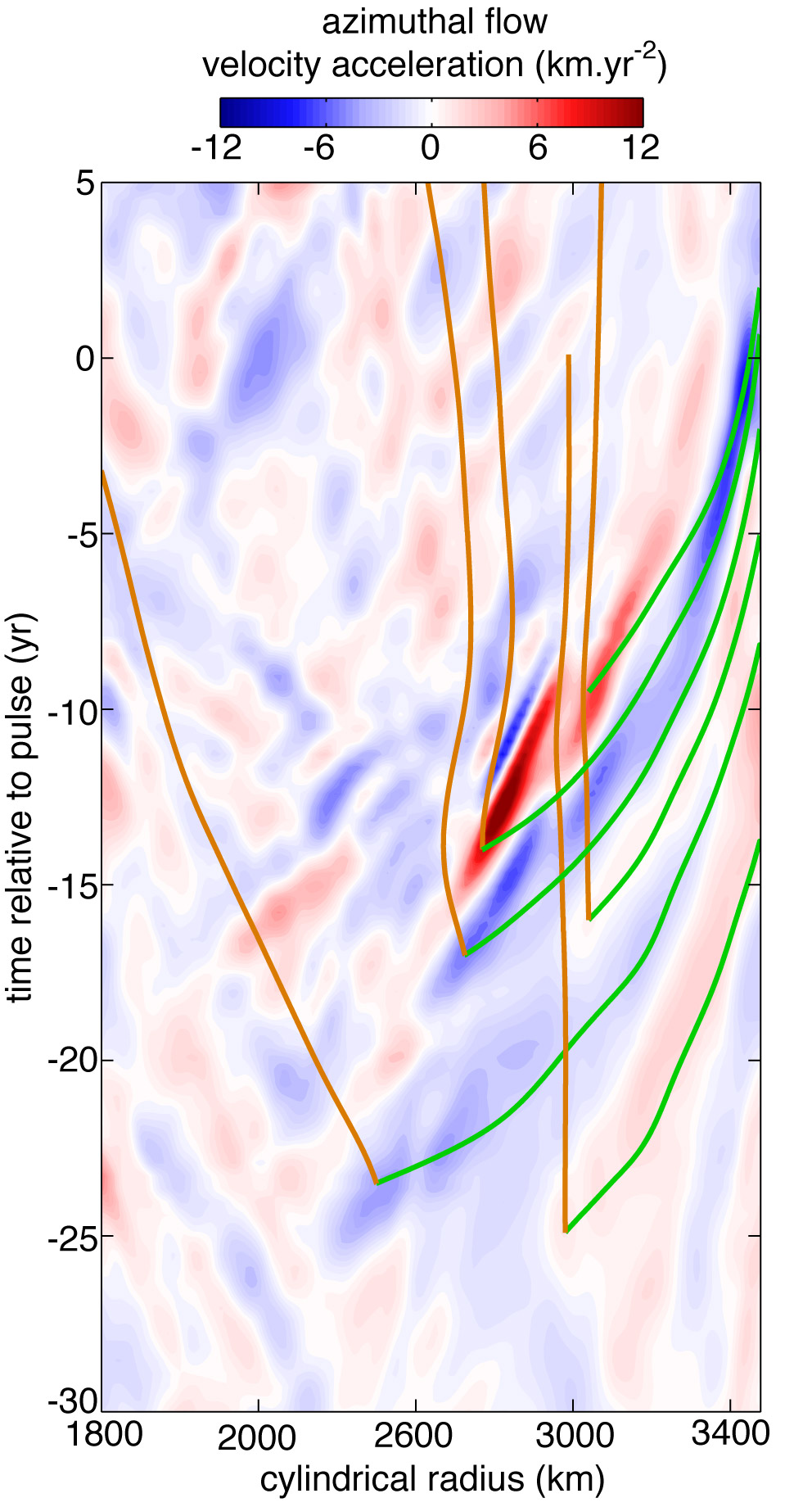}}
\caption{\label{hovdvp}Time-\rev{cylindrical} radius plot of the \rev{column-averaged} azimuthal flow acceleration \rev{$\partial \left< \vc{u}\cdot\vc{e}_{\varphi} \right>/\partial t$} in the Midpath model (blue is westwards), \rev{evaluated} at the analysis longitude defined in Fig. \ref{planforms}. The time origin corresponds to the equatorial SA pulse event located by an arrow in Fig. \ref{pathevol}d (right panel) and imaged in Fig. \ref{pulseSA}b. Green \rev{curves} represent the theoretical propagation tracks of waves at the \rev{column-averaged Alfv\'en speed $\left<(\vecB\cdot\vc{e}_{s})^{2}/\rho\mu\right>^{1/2}$ evaluated at the analysis longitude. Brown curves represent the tracks corresponding to material advection at the column-averaged cylindrical radial fluid velocity $\left<\vecu\cdot\vc{e}_s\right>$ evaluated at the analysis longitude.}}
\end{figure}

Following the previous results, sources for significant SA signatures such as the pulses observed in the Midpath model need to be searched for in non-axisymmetric flow accelerations. A few years before the occurrence of SA pulses, we can typically observe spatially localised, alternating patterns of azimuthal flow acceleration (see arrows in Fig. \ref{planforms}a) propagating in the cylindrical radial direction from the tip of convective plumes (Fig. \ref{planforms}c). These patterns are elongated in the azimuthal direction but limited in their longitudinal extent, and present an axially-invariant, columnar shape in meridional cuts (Fig. \ref{planforms}b). At cylindrical radii above 2900 km, the radial convective flow is weak (Fig. \ref{planforms}d) as the buoyancy profiles approaches the neutral state prescribed at the core-mantle boundary. Yet \rev{Fig. \ref{hovdvp} shows that these alternating columnar patterns, best seen in time-cylindrical radius maps of $\partial \left< \vc{u}\right>/\partial t\cdot\vc{e}_{\varphi}$ evaluated at a given longitude}, propagate outwards at velocities that generally match the \rev{local columnar Alfvén velocity $\left<(\vecB\cdot\vc{e}_{s})^{2}/\rho\mu\right>^{1/2}$}, and exceed (or are even reversed compared to) the \rev{columnar velocity $\left<\vecu\cdot\vc{e}_s\right>$} of material advection. \rev{Note that the slightly spiralled structure of the patterns (Fig. \ref{planforms}a) could suggest the presence of Rossby, or magneto-Rossby waves \citep[e.g.][]{Hori2015,Hori2017}, but this cannot be the case because we observe propagation in the cylindrical radial direction while Rossby waves propagate in the azimuthal direction.}

\rev{Local propagation at the Alfvén velocity in the absence of material advection indicates the presence of magneto-inertial transverse wave motion at the non-axisym\-metric level. This is rather unexpected because unlike axisymmetric torsional Alfvén waves, the present waves cannot be entirely geostrophic and should be at least locally influenced by the Coriolis force. With this force strongly dominating inertia in the rapidly rotating regime (Fig. \ref{forcebal}a), magneto-Coriolis or inertial-Coriolis waves are generally preferred over magneto-inertial Alfvén waves \citep{Finlay2010}. The influence of the Coriolis force can however be greatly mitigated by having axially columnar, transverse wave motion where the rotation vector and wave vector are orthogonal \citep{Bardsley2016}, as is presently the case. We should nevertheless clarify the mechanism that maintains the non-axisymmetric structure of the waves, and particularly the process through which the Coriolis force associated with their finite longitudinal extent can be balanced.}

\begin{figure}
\centerline{\includegraphics[width=12cm]{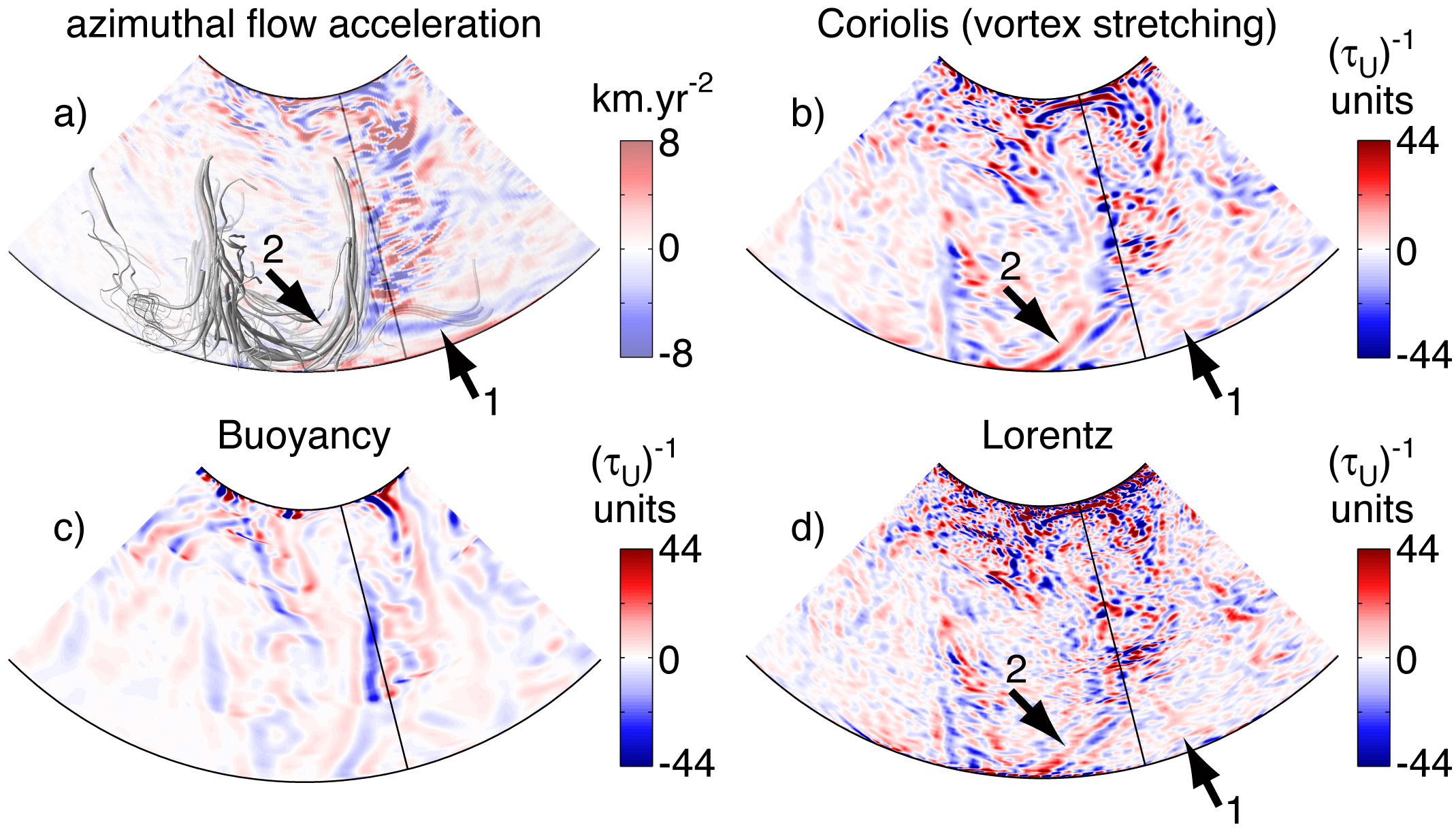}}
\caption{\label{vorticities}Partial equatorial planforms obtained  8.7 years before the equatorial SA pulse located by an arrow in Fig. \ref{pathevol}d (right panel) and imaged in Fig. \ref{pulseSA}b. a: Semi-transparent planform of the azimuthal flow acceleration (detail of Fig. \ref{planforms}a), presented together with a three-dimensional rendering of local magnetic field lines (gray, thickness proportional to the local magnetic field amplitude). b-d: Planforms of the Coriolis, buoyancy and Lorentz \rev{contributions to the first-order curled force balance (equation \ref{curlbal})}. Arrows 1 and 2 in (a,b,d) respectively mark the location of the quasi-geostrophic Alfv\'en (QGA) wavefronts visible in (a), and the location where the QGA wavefront is interrupted by magnetic field inhomogeneities. The black line in a-d marks the analysis longitude.}
\end{figure}

\begin{figure}
\centerline{\includegraphics[width=8cm]{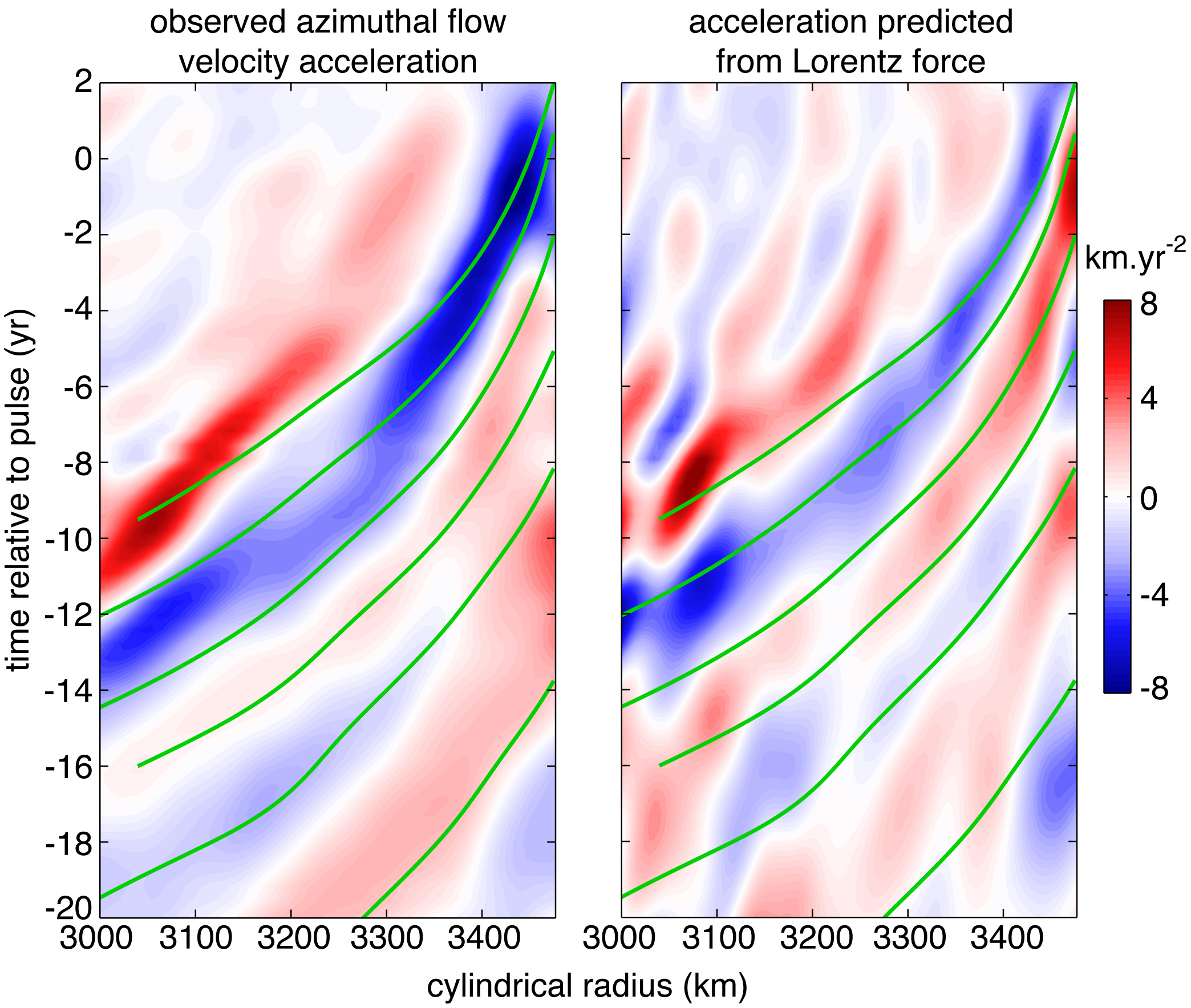}}
\caption{\label{wavebal}\rev{Time-cylindrical radius plots of the column-averaged azimuthal flow acceleration at the analysis longitude (left panel, detail of Fig. \ref{hovdvp}, left-hand-side of equation \ref{approxbal}), and of the column-averaged Lorentz force associated with the wave (right-hand-side of equation \ref{approxbal}). The green curves are the theoretical Alfvén propagation tracks from Fig. \ref{hovdvp}.}}
\end{figure}

\rev{To this end, we examine the force balance (equation 2.1 in A17) in a Midpath model snapshot, more specifically the axial component of its curl that removes the zeroth-order geostrophic balance (Fig. \ref{forcebal}a) and reveals the first-order MAC balance, written as (in dimensional form):
\begin{equation}
-2\ddp{\vecu}{z}\cdot \vc{e}_{z} \approx \dfrac{1}{\rho\Omega} \nabla\times \left(Cg_{o}\vc{r}/r_{o}\right)\cdot \vc{e}_{z} + \dfrac{1}{\rho\mu\Omega} \nabla\times \left(\left(\nabla\times\vecB\right)\times \vecB\right)\cdot \vc{e}_{z}.\label{curlbal}
\end{equation}
Here $\vc{r}$ is the radius vector and $\vc{e}_{z}$ is the unit vector parallel to the rotation axis $\vc{\Omega}$. The left-hand-side of (\ref{curlbal}) is the contribution from the Coriolis force or vortex stretching, and the terms in the right-hand side are the buoyancy and Lorentz force contributions. If fluid flow in the cylindrical radial direction can be neglected (as in the upper 500 km of the outer core, Fig. \ref{planforms}d), mass conservation also leads to
\begin{equation}
-\ddp{\vecu}{z}\cdot \vc{e}_{z} \approx \dfrac{1}{s} \ddp{\vecu}{\varphi} \cdot \vc{e}_{\varphi},
\end{equation}
meaning that the vortex stretching term in (\ref{curlbal}) evaluates the azimuthal variability of azimuthal flow structures. The contributions to equation (\ref{curlbal}) are individually represented in Fig. \ref{vorticities}b-d in units of $1/\tau_{U}$, this latter unit naturally representing the typical amount above which longitudinal disruption of azimuthal flow structures starts to take place. Using this interpretation, it is then logical that the waves (see wavefront at location marked by arrow 1 in Fig. \ref{vorticities}a) are generally found in regions where the vortex stretching is relatively weak (see region 1 in Fig. \ref{vorticities}b) and that the wavefront edge (arrow 2 in Fig.  \ref{vorticities}a,b) corresponds to localised vortex stretching values largely exceeding $1/\tau_{U}$. The possibility of wavefronts with a limited azimuthal extent then ultimately depends on whether the Coriolis force associated with this stretching can be balanced. At location 2 the balance is essentially provided by the Lorentz contribution (Fig. \ref{vorticities}d) since buoyancy does only marginally participate to the force balance in the upper outer core  (Fig. \ref{vorticities}c). This Lorentz contribution exists because of the spatial heterogeneity of the magnetic field (see field line pattern in Fig. \ref{vorticities}a at location 2). In summary, despite the constraints set by rapid rotation, quasi-geostrophic waves of limited azimuthal extent are possible within magnetically homogeneous regions of the fluid, while magnetic heterogeneity balances the rotational constraints at the edges of these regions. 

Turning now to the force balance at second order (see Fig. \ref{forcebal}a), the waves can be understood within the same theoretical framework as torsional Alfvén waves \citep[e.g.][]{Finlay2010,Jault2015treatise} if the following magneto-inertial equilibrium holds at a local level (rather than for the azimuthal averages that are usually invoked for torsional waves):
\begin{equation}
\dfrac{\partial} {\partial t} \left< \vecu\cdot\vc{e}_{\varphi} \right> \approx \dfrac{1}{\rho\mu} \left<\left(\nabla\times\vecB_{\varphi}^{\mathrm{fast}}\right)\times\vecB^{\mathrm{slow}} \right>\cdot \vc{e}_{\varphi} \label{approxbal}
\end{equation}
Here $\vecB_{\varphi}^{\mathrm{fast}}$ is the rapidly-evolving azimuthal vector component of the magnetic field that carries the perturbation induced by the wave at a time scale of order $\tau_{A}$, while $\vecB^{\mathrm{slow}}$ is the background magnetic field slowly evolving at time scale $\tau_{U}\gg \tau_{A}$. Checking the validity of equation (\ref{approxbal}) is difficult because it is not straightforward to precisely separate $\vecB_{\varphi}^{\mathrm{fast}}$ and $\vecB^{\mathrm{slow}}$ in our system. As a rough approximation, we may define $\vecB^{\mathrm{slow}}$ as the time-averaged magnetic field over the time interval spanned by Fig. \ref{hovdvp}, and $\vecB_{\varphi}^{\mathrm{fast}}$ as the azimuthal magnetic field after removal of a running time average with a window of 6 yr that matches the wave period. Fig. \ref{wavebal} shows good consistency between the wave acceleration observed within region 1 and the Lorentz acceleration term predicted from (\ref{approxbal}). Taken together, the consideration of the first-order (Fig. \ref{vorticities}) and second-order (Fig. \ref{wavebal}) force balances finally indicate that the waves observed here may be identified as quasi-geostrophic Alfvén (QGA) waves within a magnetically heterogeneous fluid.}

The configuration of the background magnetic field (field lines in Fig. \ref{vorticities}a) is essentially static during the SA pulse event, as it is shaped by the slower convective motions \rev{at time scale $\tau_{U}$}. The QGA wavefronts are both guided radially and bound laterally by this field line configuration, until they reach the core-mantle boundary at equatorial positions (see Fig. \ref{hovdvp} at time 0 yr) and provide the azimuthal, non-axisymmetric and toroidal flow accelerations that have been shown to be at the origin of the SA pulse (Fig \ref{pulseSA}b). In a system where the ratio of Alfvén to convective velocities is high (low Alfvén number), especially close to the outer boundary where the buoyancy profile approaches neutrality, the QGA waves are therefore the main carriers that transmit the kinetic energy of convective plumes to the surface to create events of significant SA. 

Similarly to the observations of \cite{Gillet2010} and simulations of \cite{Schaeffer2017}, \rev{the reflection of} torsional and QGA waves at the core-mantle boundary \rev{is elusive in Figs. \ref{torsionalwaves},\ref{hovdvp},\ref{wavebal}}. The theoretical reflection \rev{coefficient at a plane and rigid wall for Alfvén waves within a uniform magnetic field} is $R=(1-Q-\sqrt{Pm})/(1+Q+\sqrt{Pm})$, \rev{a formula that also describes well cases with nearly uniform magnetic fields in spherical geometry \citep{Schaeffer2016,Gillet2017}.} Here the contribution in $\sqrt{Pm}$ represents absorption within the electrically conducting fluid, while $Q$ is the quality factor associated with reflection at the electrically conducting boundary. Using our definitions, this quality factor may be expressed as
\begin{equation}
Q=\sqrt{\dfrac{\mu}{\rho}}\Delta \sigma_{m}B_{r}(r_{o})\approx S \dfrac{\Delta \sigma_{m}}{D \sigma_{c}} \dfrac{B_r (r_{o})}{B}.
\end{equation}
This expression involves the Lundquist number $S$, deep mantle electrical \rev{conductance $\Delta\sigma_{m}$}, and r.m.s. radial magnetic field at the core-mantle boundary $B_r (r_{o})$. The values in use for \rev{the Midpath model ($S=9100$, $\Delta \sigma_{m}/D \sigma_{c}=\te{-4}$ and $B_{r}(r_{o})/B \approx 1/8$)} lead to $Q\approx \rev{0.1}$ and $R\approx \rev{0.5}$. \rev{At these values of $Q$ and $R$, \cite{Gillet2017} have shown that the reflected wave pattern cannot be observed in a straightforward manner because it is obscured by the more intense incoming waves.} 

\section{\label{discu}Discussion.}

\rev{\subsection{Dynamic variability versus kinematic invariance along the path.}}
Our analysis of \rev{magnetic field kinematics and dynamics (the magnetic secular variation and acceleration)} in numerical dynamo simulations \rev{is linked to} the properties of the parameter space path along which these dynamos reside. It is useful to first recall some of these properties and put them in perspective with the present results. In A17, the concept of kinematic invariance has been introduced, based on the large-scale morphological stability of the velocity, magnetic and density anomaly fields along the path. Here we have seen several examples and confirmations of this concept, for instance in the invariance of the r.m.s. secular variation pattern at the core-mantle boundary (Fig. \ref{pathevol}), of the master secular variation time scale $\tau_\mathrm{SV}^{1}$ (Fig. \ref{tauSVSA}, Table \ref{SVSA}), of the axisymmetric zonal flow pattern (Fig. \ref{2D3D}). The analysis of the secular acceleration signature in the temporal domain (Fig. \ref{psdrotdom}) also suggests that the slow dynamics that occurs at time scales longer than $\tau_{U}$ is also invariant along the path. This part may be seen as corresponding to the larger spatial scales of the dynamo system, where the dynamics is mainly prescribed by a thermal wind balance between the Coriolis force, pressure gradient and buoyancy forces (Fig. \ref{forcebal}). On such long temporal and large spatial scales, the rotational constraint is already sufficiently enforced at the start of the path, and magnetic control is subdominant. This therefore provides a justification for the use of \rev{start-of-path} models and thermal wind dynamics to study and forecast the long-term evolution of the geodynamo \citep[e.g.][]{Aubert2015}. 

In contrast to these invariant properties, we have seen that the secular acceleration signal probes a dynamical structure at time scales shorter than $\tau_{U}$ that significantly changes along the path (Figs. \ref{pathevol},\ref{psdrotdom}). We have linked these changes in the rapid dynamics to the gradual enforcement of rotational dominance, and strong magnetic control (or strong-field dynamo action, A17), yielding the time scale separation 
$\tau_{U}\gg \tau_{A} \gg \tau_{\Omega}$. The secular acceleration patterns and pulses (Figs. \ref{pathevol},\ref{pulseSA}) are modified to account for the corresponding changes in the underlying flow acceleration (Figs \ref{2D3D}, \ref{planforms}). The flow acceleration patterns have revealed the presence of weakly attenuated Alfvén waves, both of geostrophic (axisymmetric, Fig. \ref{torsionalwaves}) and quasi-geostrophic (non-axisymmetric, Figs. \ref{planforms},\ref{hovdvp}) nature, propagating at speeds significantly faster than convection since $A=\tau_{A}/\tau_{U} \ll 1$. The appearance of such waves naturally echoes the progressive adherence of the dynamo system to a Taylor state along the path that has been highlighted in A17. The Taylor state may indeed be viewed as the background state over which these waves can develop. The scaling analysis performed in A17 has demonstrated that it is highly likely that the rapid rotation regime attained at the middle of the path is asymptotic, and our analysis has confirmed that the secular acceleration and underlying dynamics tend to a stable behaviour after 30\% of the path (Figs. \ref{pathevol},\ref{psdrotdom}), where all of the SA energy is contained at time scales longer than that of planetary rotation. It is interesting to note that \cite{Jault2008} has previously conjectured that the asymptotic regime for rotationally-dominated and magnetically-driven dynamics takes place at Lehnert numbers $\lambda=Ro/A=\tau_{\Omega}/\tau_{A}$ smaller than $\lambda\approx 5~\te{-3}$, a transitional value which corresponds to that obtained at 30\% of the path (Table \ref{tsratios}). We therefore conclude that this latter path position, corresponding along our path to a classical Ekman number $E=3~\te{-7}$, should be reached for numerical geodynamo simulations to give a physically realistic account of geomagnetic secular acceleration and rapid dynamics in Earth's core.

\rev{\subsection{Frequency-domain energy spectra of the geomagnetic acceleration in the models and Earth's core.}}
Models advanced along the parameter space path show that an approximately flat ($f^{0}$) or slightly bell-shaped spectral energy density range exists in temporal secular acceleration spectra (Fig. \ref{psdrotdom}), between the overturn frequency $f=1/\tau_{U}$ and frequencies on the order of $f=0.1~\ut{yr}^{-1}$. At higher frequencies the secular acceleration spectra present a $f^{-2}$ decay. Evidence for a flat intermediate range, respectively corresponding to $f^{-4}$ and $f^{-2}$ ranges in spectra of the main magnetic field and secular variation, has been previously reported in observations and numerical simulations of the geodynamo \citep[see][and references therein]{Bouligand2016}. Recent observational inferences \citep{Lesur2017} suggest that this range should extend to the shortest resolvable time scales of the internal geomagnetic field i.e. $f\approx 1~\ut{yr^{-1}}$. Advanced models such as the Midpath case obviously fall short of reproducing such an extended flat spectral range, possibly highlighting the limits of the approximations used to obtain these models. One possibility is that the use  of hyperdiffusivity damps the rapid signatures of small-scale turbulence. Comparing the frequency domain spectra of direct and large-eddy simulations does generally not support this hypothesis, but the range where both DNS and LES are feasible is restricted to moderate levels of turbulence (A17). Further analysis of the secular acceleration in fully-resolved, extreme numerical simulations \citep[e.g.][]{Schaeffer2017} is therefore necessary to investigate this issue. We note again that an extended flat secular acceleration spectral range, together with a decay less steep than, or matching the $f^{-2}$ trend, are the markers of a signal that is ill-behaved when differentiated in time once more, and hence of the possibility of geomagnetic jerks \citep{Bouligand2016}. 

\rev{\subsection{Torsional waves, quasi-geostrophic Alfvén waves and their signatures in the geomagnetic secular acceleration.}}
Though the models beyond 30\% of the path reproduce the rapid, hydromagnetic wave-driven dynamics expected at Earth's core conditions on top of convection, this new dynamics is rather subtle. Torsional waves are indeed found at a weak amplitude relatively to the convective signal (Fig. \ref{torsionalwaves}), with a subdominant contribution to the secular acceleration, both at normal and pulse epochs (Figs. \ref{normalSA}, \ref{pulseSA}). This confirms the conclusions previously obtained by \cite{Cox2016} regarding the difficulty to offer an explanation for geomagnetic jerks solely resting on torsional waves, as was initially advocated for by \cite{Bloxham2002}. Future studies should therefore attempt to highlight other possible mechanisms for these sudden changes in the geomagnetic secular acceleration. Quasi-geostrophic Alfvén waves bear a significant signature on the secular acceleration in the form of energy pulses, but this signature is intermittent (Fig. \ref{pathevol}) and does not influence integral diagnostics such as $\tau_\mathrm{SA}^{0}$ (Fig. \ref{tauSVSA}, Table. \ref{SVSA}). Together with the subdominance of torsional waves signatures in the secular acceleration, this explains why classical models found at the start of the path have already obtained values of $\tau_\mathrm{SA}^{0}$ that are very similar to geomagnetic inferences \citep{Christensen2012}. The invariance of $\tau_\mathrm{SA}^{0}$ along the path also links this diagnostic with the invariant slow convective dynamics and is therefore not indicative of rapid core dynamics. It finally suggests that despite high sensitivity to the regularisation used in geomagnetic field models, the current inference $\tau_\mathrm{SA}^{0}\approx 10~\ut{yr}$ for the geodynamo \citep{Holme2011, Christensen2012} is essentially correct. 

Among the rapid dynamics present in Earth's core \citep{Gillet2010} and also observed in advanced numerical geodynamo simulations, axisymmetric torsional waves have been repeatedly exhibited before \citep{Wicht2010,Teed2014,Teed2015,Schaeffer2017} \rev{with various simulation set-ups} and are now well understood. The non-axisymmetric, quasi-geostrophic Alfvén waves highlighted in this study (Figs. \ref{planforms}-\ref{wavebal}) are perhaps more surprising, because the classical theoretical views operate a strong dichotomy between \rev{axisymmetric and non-axisymmetric waves \citep[see e.g. a review in][]{Finlay2010}. The former are insensitive to the Coriolis force, and their rapid time dependency is governed by an Alfvénic magneto-inertial equilibrium. The latter are influenced by the Coriolis force, leading in particular to a slow magneto-Coriolis (MC) force equilibrium, with the time dependency then essentially governed by the magnetic induction equation. These views have been guided by the consideration of simple, homogeneous, or weakly heterogeneous imposed magnetic fields configurations \citep[e.g][]{Labbe2015}. Our results from self-sustained dynamos in the rapidly rotating regime highlight another possibility related to the natural production of localised magnetic field heterogeneities. Rapid Alfvénic wave motion can then locally take place within magnetically homogeneous regions of the fluid, while magnetic heterogeneity at the edges of these regions provide the slow MC equilibrium needed to balance the rotational constraints. Since the MC equilibrium is part of the more general MAC balance expected to hold in Earth's core (A17), this situation should pertain to natural conditions and the quasi-geostrophic Alfvén waves} may represent an important (albeit intermittent) dynamical component contributing to the geomagnetic secular acceleration. 

The advanced numerical dynamo simulations presented here have highlighted the existence of short, intermittent and intense pulses in the secular acceleration at the core-mantle boundary (Fig. \ref{pathevol}). Pulses near the equator are caused by non-axisymmetric azimuthal flow accelerations carried by arriving quasi-geostrophic Alfvén waves (Figs. \ref{pulseSA},\ref{planforms},\ref{hovdvp},\rev{\ref{wavebal}}). The absence of waves at the start of the path rationalises the absence of secular acceleration activity close to the equator (Fig. \ref{pathevol},\ref{2D3D}). Such waves are important carriers of flow acceleration in the equatorial plane, between the cylindrical radius at which convective plumes stall and the core-mantle boundary. \rev{They are most easily exhibited if a nearly convectively neutral region exists below the core-mantle boundary, as done in the Coupled Earth (CE) set-up by imposing neutral mass anomaly flux outer boundary conditions. It should however be possible to exhibit them in other set-ups, for instance within} a stably stratified upper outer core \citep[e.g.][]{Buffett2014}. The numerical reproduction of pulses is a potentially important result from the standpoint of observational geomagnetism, because it should establish the physical reality of secular acceleration pulses observed in geomagnetic field models within the satellite era. It is likely that geomagnetic field models do not image these events in the full extent of their spatial and temporal structure (Fig. \ref{hires}), but the good coherence observed between simulated secular acceleration maps at low and high spatial resolutions \citep[as was previously suggested concerning the secular variation,][]{Holme2011} gives support to the possibility of a meaningful physical analysis at the currently available resolution. This hence opens interesting perspectives for future analysis of the dynamical origin of geomagnetic jerks, which have been linked to these secular acceleration pulses \citep{Chulliat2010,Chulliat2014}. \rev{One avenue for forthcoming work will be to perform a detailed study of the temporal numerical model behaviour in the vicinity of secular acceleration pulses, to better understand the observed quasi-periodic recurrence of jerks within the satellite era.} The infrequent secular acceleration pulses occurring at high latitudes may also be interesting from a geophysical standpoint, but require further analysis given the small-scale flow and magnetic field interactions that appear to create them. It would be interesting to put them in perspective with recent observations involving significant secular acceleration events at high latitude \citep{Chulliat2010b,Livermore2017}. In particular, expulsion of magnetic flux is unlikely to represent a viable mechanism at rapidly rotating conditions (Fig. \ref{pulseSA}), as shown by the disappearance of diffusive secular acceleration pulses after 30\% of the parameter space path.

\section*{Acknowledgements}
JA wishes to thank Christopher C. Finlay for insightful discussions and suggestions \rev{as well as Nathanaël Schaeffer and an anonymous referee for useful review comments}, and acknowledges support from the Fondation Del Duca of Institut de France (2017 Research Grant). Numerical computations were performed at S-CAPAD, IPGP and using HPC resources from GENCI-IDRIS and GENCI-CINES (Grant 2016-A0020402122). This is IPGP contribution 3929. 

\bibliographystyle{gji}
\bibliography{Biblio}

\end{document}